\PassOptionsToPackage{dvipsnames}{xcolor}
\documentclass[times]{aastex631}

\usepackage{amsmath}
\usepackage{CJK}
\usepackage{bm}
\usepackage[normalem]{ulem}
\usepackage{multirow}

\makeatletter
\newcommand*{\rom}[1]{\expandafter\@slowromancap\romannumeral #1@}
\makeatother

\usepackage{dcolumn}
\newcolumntype{q}[1]{D{.}{.}{#1}}

\usepackage{newunicodechar,graphicx}
\DeclareRobustCommand{\okina}{%
  \raisebox{\dimexpr\fontcharht\font`A-\height}{%
    \scalebox{0.8}{`}%
  }%
}

\newcommand{\chg}[1]{#1}

\begin{document}

\begin{CJK*}{UTF8}{gbsn}
\title{Towards a Robust Estimate of the Solar Photospheric Poynting Flux and Helicity Flux}

\author[0000-0002-7290-0863]{Jiayi Liu (刘嘉奕)}
\altaffiliation{\textit{DKIST} Ambassador}
\affiliation{Institute for Astronomy, University of Hawai$\okina$i at M\={a}noa, 2680 Woodlawn Dr., Honolulu, HI 96822, USA}
\email{jiayiliu@hawaii.edu}

\author[0000-0003-4043-616X]{Xudong Sun (孙旭东)}
\affiliation{Institute for Astronomy, University of Hawai$\okina$i at M\={a}noa, 34 Ohia Ku Street, Pukalani, HI 96768, USA}

\author[0000-0003-1522-4632]{Peter W. Schuck}
\affiliation{Heliophysics Science Division, NASA Goddard Space Flight Center, 8800 Greenbelt Rd., Greenbelt, MD 20771, USA}

\author[0000-0002-1198-5138]{Lars K. S. Daldorff}
\affiliation{The Catholic University of America, 620 Michigan Ave., N.E. Washington, DC 20064 USA}
\begin{abstract}

The observed solar photospheric magnetic fields and Doppler velocities are frequently used to quantify the Poynting flux and helicity flux. Multiple methods have been developed for this purpose, but their estimates of the Poynting flux and helicity flux often differ from one another. Here we study the performance of three widely used methods on NOAA active region 12673: ``PTD-Doppler-FLCT Ideal'' (PDFI), ``Differential Affine Velocity Estimator for Vector Magnetograms'' (DAVE4VM), and an extension of the latter with Doppler velocity constraint (DAVE4VMwDV). We find that the values of the accumulated energy and helicity differ significantly between the three methods, even in signs. Using the Helmholtz-Hodge decomposition, we show that Doppler velocity can contribute significantly to the Poynting flux and helicity flux through the non-inductive (curl-free) electric field. The different, ad hoc treatments of the Doppler and transverse velocities in three methods are directly responsible for the discrepancies. We discuss the desired future observations that can better constrain these methods.

\end{abstract}



\section{Introduction} \label{sec:intro}

Solar activities like flares and coronal mass ejections (CMEs) are powered by magnetic energy. The detailed dynamics are additionally constrained by relative magnetic helicity, which is a \chg{measure} of the magnetic field's connectivity and complexity \citep{Beger_1984,Nindos2004,Green2018}. An accurate estimate of magnetic energy and helicity in a coronal volume is thus important to our understanding of solar \chg{activity}. Nevertheless, the task is challenging due to the difficulty in measuring coronal magnetic fields. The magnetic field itself is generally weak, evolves rapidly, and the observations will suffer from the effect of line-of-sight (LOS) integration. Alternatively, we may estimate the magnetic energy (Poynting) flux and helicity flux as they pass through the photosphere and accumulate in the corona \chg{\citep{Welsch2015,Tilipman}}, as the photospheric observations are routinely available.

Quantifying Poynting flux and helicity flux requires the knowledge of photospheric vector velocity field $\bm{v}$ or electric field $\bm{E}$, which can not be obtained from remote sensing observations in a straightforward fashion. Multiple advanced methods have been developed to infer $\bm{v}$ and $\bm{E}$ from a time sequence of observations, for example, maps of magnetic field $\bm{B}$ (\chg{magnetograms}). For example, the local correlation tracking \citep[LCT;][]{LCT} algorithms \chg{estimate} the velocity by finding the displacement of features between two successive images that maximizes their correlation. An updated version, Fourier LCT \citep[FLCT;][]{welsch2004, FLCT}, finds the correlation function in Fourier space, which provides a faster and more efficient estimate of velocity. Nevertheless, as suggested by \cite{schuck2006}, velocity inferred by the LCT-based methods only satisfies the advection equation, which does not govern the evolution of the magnetic field. Improvement can be achieved by incorporating the ideal induction equation that describes the physical relation between velocity $\bm{v}$ and magnetic field $\bm{B}$, especially its vertical component \citep[e.g.,][]{Kusano2002,welsch2004,Longcope2004,schuck}: 
\begin{equation}
 \frac{\partial B_z}{\partial t} = \nabla_h \cdot (\bm{v}_h B_z - v_z\bm{B}_h),
 \label{equ:induction}
\end{equation}
where subscript $h$ denotes the horizontal direction and $z$ denotes the vertical direction. The operator $\nabla_h$ acts on the horizontal components alone. Similarly, the electric field can be inferred with the combination of induction equation and Faraday's law, through techniques such as the poloidal-toroidal decomposition \citep[PTD;][]{fisher2010} of the magnetic field. The results can be further improved by incorporating the observed Doppler velocity as an \chg{additional} constraint \citep{Fisher_2012}.

In \cite{welsch2007}, these velocity and electric field inversion methods have been tested and validated on the output from anelastic MHD \citep[ANMHD;][]{Fan1999,Abbett2000,Abbett2004} simulation. They found that all the methods have comparable and reasonably \chg{good}  performance on recovering the velocity field, Poynting and \chg{helicity} flux on ANMHD simulation. The updated velocity inversion method, Differential Affine Velocity Estimator for Vector Magnetogram \citep[DAVE4VM, ][]{schuck}, further \chg{improves} the performance and can reproduce $75\%$ and $95\%$ of the total Poynting and helicity flux, respectively. The electric field inversion method PTD and its update that incorporates Doppler and transverse velocities, ``PTD-Doppler-FLCT Ideal" \citep[PDFI,][]{PDFI_2014}, also have good performance and can infer less than $1\%$ error in the total Poynting flux and a $10\%$ error in the helicity flux. \chg{\cite{Afanasyev2021} further validated PDFI on a more complex simulation with magnetic flux eruption code \citep[MFE, ][]{Fan2017} and found that the accuracy of PDFI can decrease during the shearing evolution phase of an active region.}

Despite the wide application of these methods \textit{individually} on observational data, the consistency of results across multiple methods \chg{is} not well tested. In a few relevant studies, the estimated energy and helicity fluxes generally do not agree well. For example, \cite{PDFI_2015} found the difference of energy estimated by the PDFI method and the DAVE4VM on NOAA active region (AR) 11158 is about $15\%$. \cite{Lumme_2019} further studied the effect of cadence, and found that the PDFI is less sensitive to varying cadence than DAVE4VM. One explanation is that DAVE4VM is designed to comply with the induction equation only in a statistical sense (see Section~\ref{sec:discussion}).

This paper aims to investigate the performance of three widely used algorithms, PDFI, DAVE4VM, and DAVE4VMwDV (DAVE4VM's extension with Doppler velocity constraint) on observations. Specifically, we will apply these algorithms to AR 12673 observed by Helioseismic and Magnetic Imager \citep[HMI;][]{schou2012} on board the Solar Dynamics Observatory \citep[SDO;][]{SDO} and infer the Poynting flux and helicity flux. Our goals are three-fold. Firstly, we show that the three methods yield markedly different results in the presence of large Doppler velocity, even with opposite signs. Secondly, we provide a theoretical explanation to the discrepancy by isolating the contributions from the inductive and non-inductive electric fields. Finally, we propose a post-processing procedure for DAVE4VMwDV that can reduce the discrepancies. 

The rest of the paper is organized as follows. In Section~\ref{sec:methods}, we describe the observational data and the analysis methods. In Section~\ref{sec:result}, we present the results on estimated Poynting flux and helicity flux, as well as the contributions from inductive and non-inductive electric field. In Section~\ref{sec:discussion}, we discuss the possible reasons for the discrepancies in Poynting flux and helicity flux, and present a new post-processing procedure. Finally, we draw our conclusions in Section~\ref{sec:conclusion}. 


\section{Observation and Data Analysis} \label{sec:methods}
\subsection{\textit{SDO/HMI} Observation}

HMI measures the full-disk Stokes profiles at six wavelengths across the Fe I line at 617.3 nm with a 12 minutes cadence and a $0.5 ''$ plate scale. These Stokes profiles are used to invert vector magnetic field $\bm{B}$ and Doppler velocity $v_l$ with Very Fast Inversion of the Stokes Algorithm \citep[VFISV;][]{Borrero2012}. The $180^\circ$ ambiguity in the horizontal magnetic field is then resolved by Minimum Energy method \citep[ME0;][]{ME0,leka}. 

In this work, we use vector magnetograms and Dopplergrams prepared by the Coronal Global Evolutionary Model  \citep[CGEM,][]{cgem} consortium. These are automatically extracted maps for ARs, which include $\bm{B}$, $v_l$, and the LOS unit vector $\hat{\bm{\eta}}$. The maps are in a Plate Carr\'{e}e projection with equally sized longitude $\theta$ and latitude $\phi$ grid $\Delta \theta = \Delta \phi = 0.03^\circ$. The contribution from differential rotation, the relative movement of SDO with respect to the Sun, and the convective blueshift bias to Doppler velocity are further calibrated as described in \cite{welsch2013}. We further exclude frames with non-optimal quality.

This work uses the HMI observation for NOAA AR 12673 in the period starting from 2017 September 2, 00:00 UT to 2017 September 8, 05:00 UT, which covers the evolution of this AR from S08E11 to S09W70. The field of view is $20.6^\circ \times 13.4^\circ$. This active region produced four X-class flares during its passage across the solar disk \citep{12673_flare}. It is well-studied, including its magnetic configuration, flow field, and eruptivity \citep{12673flow,sun2017,Wang_2018,Vemareddy2019,Moraitis_2019}.

\begin{figure}[t!]
  \centering
  \includegraphics[width=0.47\textwidth]{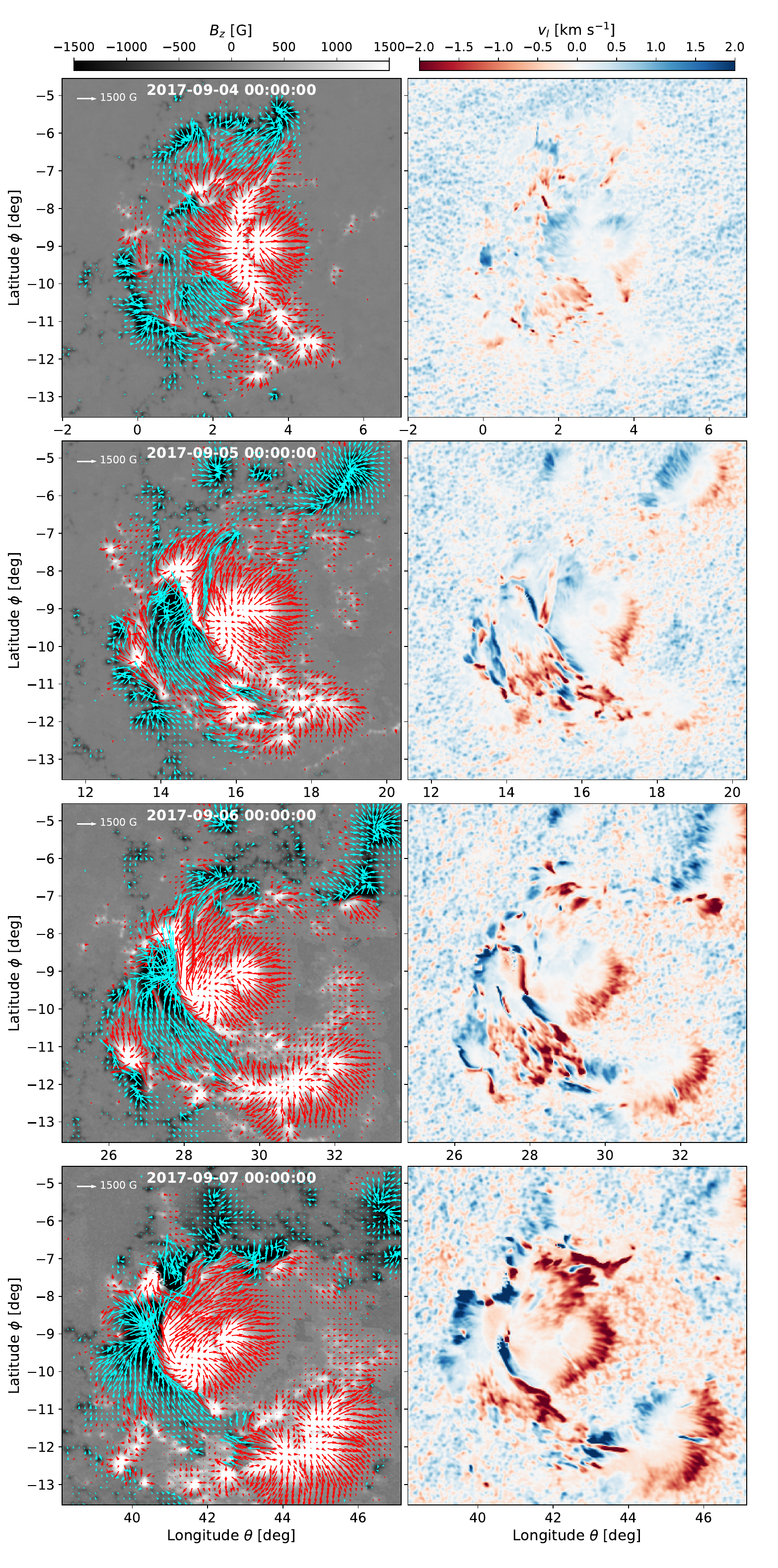}
  \caption{Overview of the evolution of the \chg{magnetic field} and Doppler velocity NOAA AR 12673 from HMI observation on four different \chg{times}. Left column: Vector magnetic field maps. The background gray map shows the vertical magnetic field $B_z$. The red (cyan) arrows \chg{show} the horizontal field vectors in positive- (negative-) $B_z$ regions with \chg{magnetic} field strength $\lvert B \rvert > 250$ G. Right column: Doppler velocity maps. Blue (red) regions have positive (negative) $v_l$. They represent blueshifted (redshifted) regions, i.e, the flow is toward (away from) the observer.} 
  \label{fig:AR12673}
  \vspace{2mm}
\end{figure}

\begin{figure}[t!]
  \centering
  \includegraphics[width=0.47\textwidth]{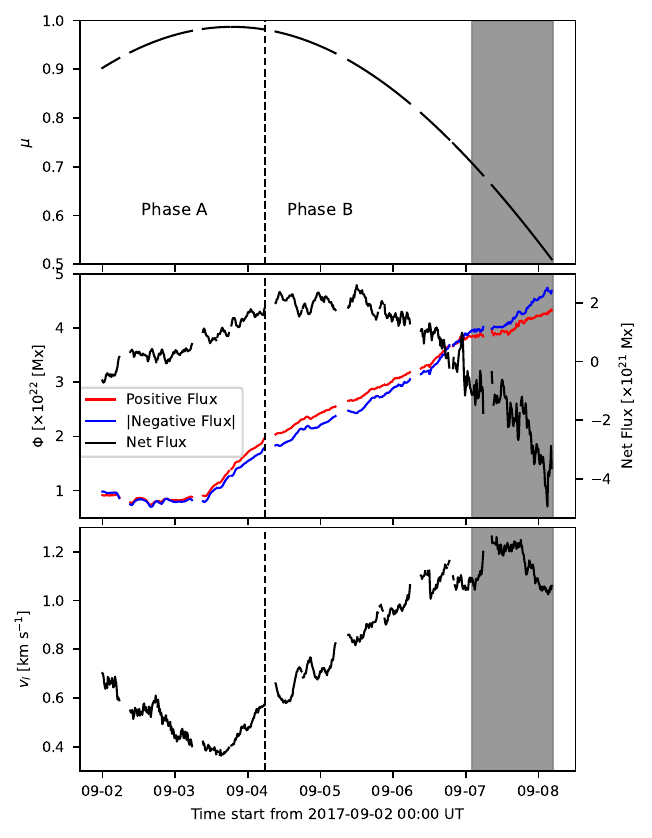}
  \caption{The time evolution of the location, magnetic flux and Doppler velocity for NOAA AR 12673. Top: The time evolution of $\mu$ (cosine of the heliocentric angle) for the centroid of the AR. Middle: The time evolution of magnetic flux $\Phi$. Red, blue, and black curves refer to positive, absolute negative, and net magnetic fluxes. The shadow region marks the time when the center of the active region is $45^\circ$ away from the disk center. Bottom: The time evolution of the $90$ percentile of Doppler velocity $v_l$ in \chg{regions} with magnetic field strength $\lvert B \rvert > 250$ G. Several data gaps are present due to eclipses.} 
  \label{fig:AR12673_flux}
  \vspace{2mm}
\end{figure}

An overview of the magnetic field and Doppler velocity evolution is presented in Figure~\ref{fig:AR12673}. The time profiles of the AR location and its magnetic flux are shown in Figure~\ref{fig:AR12673_flux}. The time profile of the $90$ percentile of the absolute Doppler velocity $v_l$ in the region with magnetic field strength $B > 250$ G is shown in the bottom panel of Figure~\ref{fig:AR12673_flux}. Based on the evolution of the magnetic field and the behaviors of the Doppler velocity, we roughly divide the time range studied in this work into two phases. In phase A, starting with a pre-existing sunspot with positive polarity, AR 12673 experienced a rapid flux emergence till September 4th, 05:48 UT. Multiple bipole pairs then emerged and started to interact. The Doppler velocity in this period is relatively small. In phase B, after the rapid emergence, the \chg{collisional shearing \citep{Chintzoglou2019}} between dipoles formed a compact $\delta$ sunspot group. Fast Doppler flows started to appear on the limb-side of the sunspots due to Evershed flow \citep{EvershedFlow}. They also appeared near the polarity inversion line (PIL) with the value up to $3.2$ km s$^{-1}$ for about $20$ hours \citep{Liu2023}. As the active region moved toward the west limb, the magnitude of Doppler velocity kept increasing and reached $7$ km s$^{-1}$, partly due to the projection effect, which is beyond the limit of HMI measurement \citep{Centeno2014}.


\subsection{Inference of Velocity Field}
We infer the velocity field from the observed magnetograms with DAVE4VM \citep{schuck}, a widely used local velocity estimator. It finds the velocity field within a subregion that minimizes the L2 form of the normal component of the induction equation. Its loss function is
\begin{equation}
 L_\text{DAVE4VM} = \sum_N \omega_{\partial_t B_z} \left\{ \frac{\partial B_z}{\partial t} + \bm\nabla _h \cdot (B_z \bm{v}_h+\bm{B}_hv_z)\right\}^2,
 \label{l1}
\end{equation}
where $\bm{B}$ is the input vector magnetogram, and $\bm{v}$ is the output velocity. \chg{The change of vertical magnetic field $\frac{\partial B_z}{\partial t}$ is calculated with vertical magnetic field $B_z$ from two successive frames with cadence $\Delta t$
\begin{equation}
 \frac{\partial B_z}{\partial t} = \frac{B_z (t + \Delta t) - B_z(t)}{\Delta t}
\end{equation}}
The subscript $N$ denotes a subregion with a size of $N \times N$ in pixels, and $\omega_{\partial_t B_z}$ refers to the weighting function. In this work, it is chosen as
\begin{equation}
  \omega_{\partial_t B_z} = \sigma_{\partial_t B_z}^{-2}
\end{equation}
where $\sigma_{\partial_t B_z}$ is the uncertainty of variable $\partial_t B_z$ \chg{and is calculated with uncertainty of vertical magnetic field $\sigma_{B_z}$
\begin{equation}
  \sigma_{\partial_t B_z} = \frac{\sqrt{\sigma_{B_z}^2 (t + \Delta t) + \sigma_{B_z}^2(t)}}{\Delta t}.
\end{equation}}

We also infer the velocity with DAVE4VMwDV, an updated version of DAVE4VM that estimates velocity with the additional constraint of Doppler velocity $v_{l}$. The modified method has the following loss function
\begin{equation}
  \begin{split}
 L &= L_\text{DAVE4VM} + \lambda L_\text{DV},  \\
 L_\text{DV} &= \sum_N \omega_{v_{l}} (\hat{\bm{\eta}}\cdot \bm{v}-v_{l})^2,
  \label{l2}
  \end{split}
\end{equation}
where $\lambda$ is a scalar multiplier that controls the importance of the Doppler loss term $L_{DV}$, and $\omega_{v_{l}}$ refers to the weighting for \chg{the residual} of Doppler velocity and in this work is chosen as
\begin{equation}
  \omega_{v_l} = \sigma_{v_l}^{-2}
\end{equation}
where $\sigma_{v_l}$ is the uncertainty of the LOS (Doppler) velocity $v_{l}$.

Both versions model $\bm{v}$ \chg{as a linear combination of low-order Legendre polynomials within a window} \citep{Legendre}:
\begin{equation}
 \bm{v}(x, y) = \sum_{i=0}^d \sum_{j=0}^{d-i}\bm{A}_{i,j} P_i(x) P_j(y),
 \label{equ:legendre}
\end{equation}
where $x$ and $y$ are the coordinate within the window, $P_i$ is the $i$-th order Legendre polynomial, $\bm{A}_{i,j}$ is the coefficient for $i$-th and $j$-th order of Legendre polynomial for $x$ and $y$ direction, and $d$ is the maximum order of Legendre polynomial, which is a free parameter that can be optimized. The coefficients $\bm{A}_{i,j}$ are solved with least square methods by minimizing $L$ in Equation~(\ref{l2}). The resulting $\bm{v}$ at the center of the window is set as the solution for the pixel where the window is centered. In practice, the degree of Legendre polynomials describing the vertical components of velocity $v_z$ can be set separately with \chg{a} free parameter $d_r$.

Following \cite{Liu2023}, we choose the following values: \chg{window size} $w = 23$, \chg{degree of Legendre polynomials for horizontal velocity} $d=5$, and \chg{degree of Legendre polynomials for vertical velocity} $d_r = 7$. We use $\lambda = 0.4$ for DAVE4VMwDV, and $\lambda = 0.0$ for DAVE4VM (no Doppler velocity constraint).

We emphasize here the \textit{local} nature of the DAVE4VM methods. The solution $\bm{v}$ is designed to minimize $L$ within a specific window, and is largely independent from the solutions of neighboring pixels based on different windows. As such, the $\bm{v}$ maps are not suited for differential operations, i.e., one should not apply the horizontal divergence operator on $B_z\bm{v}_h-v_z\bm{B}_h$ to evaluate the second term of Equation~(\ref{l1}). Rather, one should evaluate the term in individual windows, where \chg{it is} formally minimized. \chg{The current version of DAVE4VM(wDV) has the capability to implement global solutions that enforce the induction equation to the precision chosen by user and exports the inferred $\nabla \times (\bm{v} \times \bm{B}) \cdot \hat{\bm{z}}$ and $\nabla_h \cdot (\bm{v} \times \bm{B})$.}

Within each window, the spatial derivatives of the velocity $\bm{v}$ can be calculated as a sum of weighted derivatives of basis Legendre polynomials:
\begin{equation}
 \frac{\partial \bm{v}}{\partial x_k} = \sum_{i=0}^d \sum_{j=0}^{d-i}\bm{A}_{i,j} \frac{\partial P_i(x) P_j(y)}{\partial x_k},
  \label{equ:internal}
\end{equation}
where $\{x_k\} = \{x_1, x_2\} = \{x, y\}$. We dub this as the ``internal derivative'', which is used by DAVE4VM(wDV) to minimize the loss functions. \chg{Using this derivative, the $z$-component of the right-hand side of the induction equation (Equation~(\ref{equ:induction})), calculated at the center of the window, can be considered as the least-square ``fitted'' value of $\partial B_z^{\text{LS}} /\partial t$, and is provided as the output of the code.} This differs somewhat from the common practice, where $\nabla \times (\bm{v} \times \bm{B}) \cdot \hat{\bm{z}}$ is calculated by applying the curl operation on the final $\bm{v}$ and $\bm{B}$ maps. We will discuss the implications in Section~\ref{sec:discussion}.

\subsection{Inference of Electric Field}
The electric field is inferred from successive magnetograms and Dopplergrams using the PDFI \citep{PDFI_2014} method. It solves for the electric field $\bm{E}$ \chg{in staggered grid} that satisfies
\begin{equation}
 \bm{E} = \bm{E}^\text{I} - \nabla_h \psi - \nabla \psi^\text{C},
 \label{equ:E}
\end{equation}
where $\bm{E}^\text{I}$ is the so-called ``inductive'' electric field, which satisfies Faraday's law
\begin{equation}
  \nabla \times \bm{E}^\text{I} = - \frac{\partial \bm{B}}{\partial t},
  \label{equ:EI}
\end{equation}
and \chg{the term $-\nabla_h \psi - \nabla \psi^\text{C}$ is the ``non-inductive'' electric field} that is not constrained by the observed magnetograms. Here $\bm{E}^\text{I}$ is solved using the same PTD \citep{fisher2010} algorithm as in PDFI. The PDFI method then finds $\psi$ by using the observed Doppler velocity and horizontal velocity inferred by FLCT and solving three Poisson equations
\begin{equation}
  \label{equ:PDFI_pipeline}
  \begin{split}
  \psi &= \psi^\text{D} + \psi^\text{F}, \\ 
  \nabla_h^2 \psi^\text{D} &= \omega \nabla \cdot (v_l\hat{\bm{\eta}} \times \bm{B}), \\
  \nabla_h^2 \psi^\text{F} &= (1 - \omega)\nabla_h \cdot (\bm{v}_h^\text{F} \times B_z \hat{\bm{z}}).
  \end{split}
\end{equation}
Here, the superscripts $\text{D}$ and $\text{F}$ refer to terms relevant to the Doppler velocity and the FLCT output, respectively. The parameter $\omega$ is an ad hoc weighting function for pixels near the PILs where the Doppler velocity is considered important. It is based on the LoS magnetic field $B_l$ and the transverse magnetic field strength $B_t$
\begin{equation}
  \omega = \exp \left[-\frac{1}{\sigma_\text{PIL}^2} \left\lvert\frac{B_l}{B_t}\right\rvert^2 \right]
  \label{equ:PIL_weighting}
\end{equation}
where $\sigma_{\text{PIL}}$ is an adjustable parameter that controls the ``effective width'' of PIL, where the contribution of the Doppler velocity is expected to be large. That is, for a larger $\sigma_{\text{PIL}}$, the weighting $\omega$ increases, so the Doppler signal from pixels away from the PIL will contribute more. Its value may be tuned based on MHD models where the electric field is known. In this work, $\sigma_\text{PIL}$ is set as $1$, which is PDFI's default value that \chg{is} tuned with ANMHD model. We choose Dirichlet boundary conditions for these two Poisson equations:
\begin{equation}
  \begin{split}
  \psi^\text{D} &= 0, \\
  \psi^\text{F} &= 0.
  \end{split}
\end{equation}
The third term $\psi^\text{C}$ in Equation \ref{equ:E} enforces that the electric field $\bm{E}$ satisfies the ideal MHD condition $\bm{E} \cdot \bm{B} = 0$, i.e.
\begin{equation}
  \nabla \psi^\text{C} \cdot \bm{B} = (\bm{E}^\text{I} - \nabla \psi^\text{D} - \nabla \psi^\text{F})\cdot \bm{B}. 
\end{equation}
It is solved with the ``iterative" method described in \cite{fisher2010}. 


\subsection{Helmholtz-Hodge Decomposition}

The ideal electric field $\bm{E}$ (as inferred from PDFI) is related to velocity field $\bm{v}$ and magnetic field $\bm{B}$ via Ohm's law
\begin{equation}
 \bm{E} = -\bm{v} \times \bm{B}.
  \label{equ:ohm}
\end{equation}
While $\bm{E}$ can be uniquely determined from $\bm{v}$ and $\bm{B}$, the solution of $\bm{v}$, when $\bm{E}$ and $\bm{B}$ are known, is not unique. For this reason, we choose to compare the three methods in terms of $\bm{E}$. 

We perform the Helmholtz-Hodge decomposition to separate the electric field into a divergence-free term, a curl-free term, and a vertical term following \cite{schuck_2019},
\begin{equation}
  \label{equ:decomposition}
 \bm{E} = \hat{\bm{z}}\times \nabla \chi + \nabla_h \xi + \tau \hat{\bm{z}},
\end{equation}
where $\hat{\bm{z}}$ is the unit normal vector, \chg{$\chi$ and $\xi$ }are the scalers for the divergence-free and the curl-free terms, and $\tau$ is the projection of $\bm{E}$ in $\hat{\bm{z}}$ direction. For the velocity field $\bm{v}$ inferred from DAVE4VM and DAVE4VMwDV and observed $\bm{B}$, we have
\begin{equation}
  \tau = - \hat{\bm{z}}\cdot (\bm{v} \times {\bm{B}}).
\end{equation}
Furthermore, $\chi$ and $\psi$ can be uniquely determined by solving the following Poisson equations on surface $S$,
\begin{equation}
  \label{equ:solution}
  \begin{split}
  \nabla_h^2 \chi &= \nabla \times (\bm{v} \times \bm{B}) \cdot \hat{\bm{z}}, \\ 
  \nabla_h^2 \xi &= \nabla_h \cdot (\bm{v} \times {\bm{B}}).
  \end{split}
\end{equation}
We choose the boundary conditions as
\begin{equation}
  \label{equ:boundary}
  \begin{split}
 \frac{\partial \chi}{\partial n} &= 0, \quad \partial S,\\
    \xi &= 0, \quad \partial S.
  \end{split}
\end{equation}
Note that if the surface integral of $\nabla \times (\bm{v} \times \bm{B})$ is not zero, we will need to correct the boundary condition for $\chi$ with the method described in \cite{fisher2010}. These Poisson equations are solved \chg{in staggered grid} with subroutine \texttt{HSTSSP} in \texttt{FISHPACK}\chg{ \citep{Fishpack}}, which is also used for PDFI calculations. 

\chg{This decomposition has a similar formula to the electric field from PDFI. Compared to Equations~(\ref{equ:E}) and (\ref{equ:EI}), it is easy to see that the divergence-free term corresponds to the inductive term, and the curl-free term corresponds to the non-inductive term $\bm{E}^\text{NI}$, i.e.,}
\begin{equation}
  \begin{split}
 \bm{E} &= \bm{E}_h^\text{I} + \bm{E}_h^\text{NI} + \tau \hat{\bm{z}}, \\
 \bm{E}_h^\text{I} &= \hat{\bm{z}}\times \nabla \chi, \\ 
 \bm{E}_h^\text{NI} &= \nabla_h \xi .
  \end{split}
  \label{equ:de_E}
\end{equation}
\chg{For comparison, $\bm{E}^\text{I}_h$ and $\bm{E}^\text{NI}_h$ from PDFI satisfy}
\begin{equation}
  \begin{split}
 \nabla \times &\bm{E}_h^\text{I} = -\frac{\partial B_z}{\partial t}, \\ 
 \bm{E}_h^\text{NI} &= -\nabla_h \psi - \nabla_h \psi^\text{C} .
  \end{split}
  \label{equ:de_E_PDFI}
\end{equation}
The related constraints to obtain the inductive electric field $\bm{E}^\text{I}$ and non-inductive electric field $\bm{E}^\text{NI}$ from DAVE4VM(wDV) and PDFI are listed in Table \ref{tab:quantity}.

\begin{deluxetable}{ccc}[t!]
  \tablecaption{Constraints for $\bm{E}_h^\text{I}$ and $\bm{E}_h^\text{NI}$}\label{tab:quantity}
  \tablewidth{0pt}
  \tablehead{
  \colhead{\textbf{}} & \colhead{\textbf{DAVE4VM(wDV)}} & \colhead{\textbf{PDFI}}
 }
 \decimals
  \startdata
    $\bm{E}_h^\text{I}$ & $\nabla \times (\bm{v}^\text{DAVE}\times \bm{B}) \cdot \hat{\bm{z}}$ & $\partial_t B_z^{\text{Obs}}$   \\ 
    \hline 
    \multirow{3}{*}{$\bm{E}_h^\text{NI}$} & & $(1-\omega)\nabla_h \cdot (\bm{v}_h^F\times B_z \hat{\bm{z}})$ \\
 & $\nabla_h \cdot (\bm{v}^\text{DAVE}\times \bm{B})$ & $\omega\nabla \cdot (v_l \hat{\bm{\eta}}\times \bm{B})$ \\
 & & $\bm{E} \cdot \bm{B} = 0$
  \enddata
   \tablecomments{\chg{$\bm{E}_h^\text{I}$ and $\bm{E}_h^\text{NI}$ are the inductive and non-inductive electric fields, respectively.} $\bm{v}^\text{DAVE}$ represents the velocity field inferred by DAVE4VM or DAVE4VMwDV. $\bm{v}^\text{F}$ represents the velocity field inferred by FLCT. $v_l$ represents the observed Doppler velocity. \chg{The superscript ``Obs'' in $\partial_t B_z^{\text{Obs}}$ emphasizes that the observed magnetograms are used directly to assess the temporal derivative.}}
  \end{deluxetable}


\begin{figure*}[t!]
  \centering
  \includegraphics[width=0.95\textwidth]{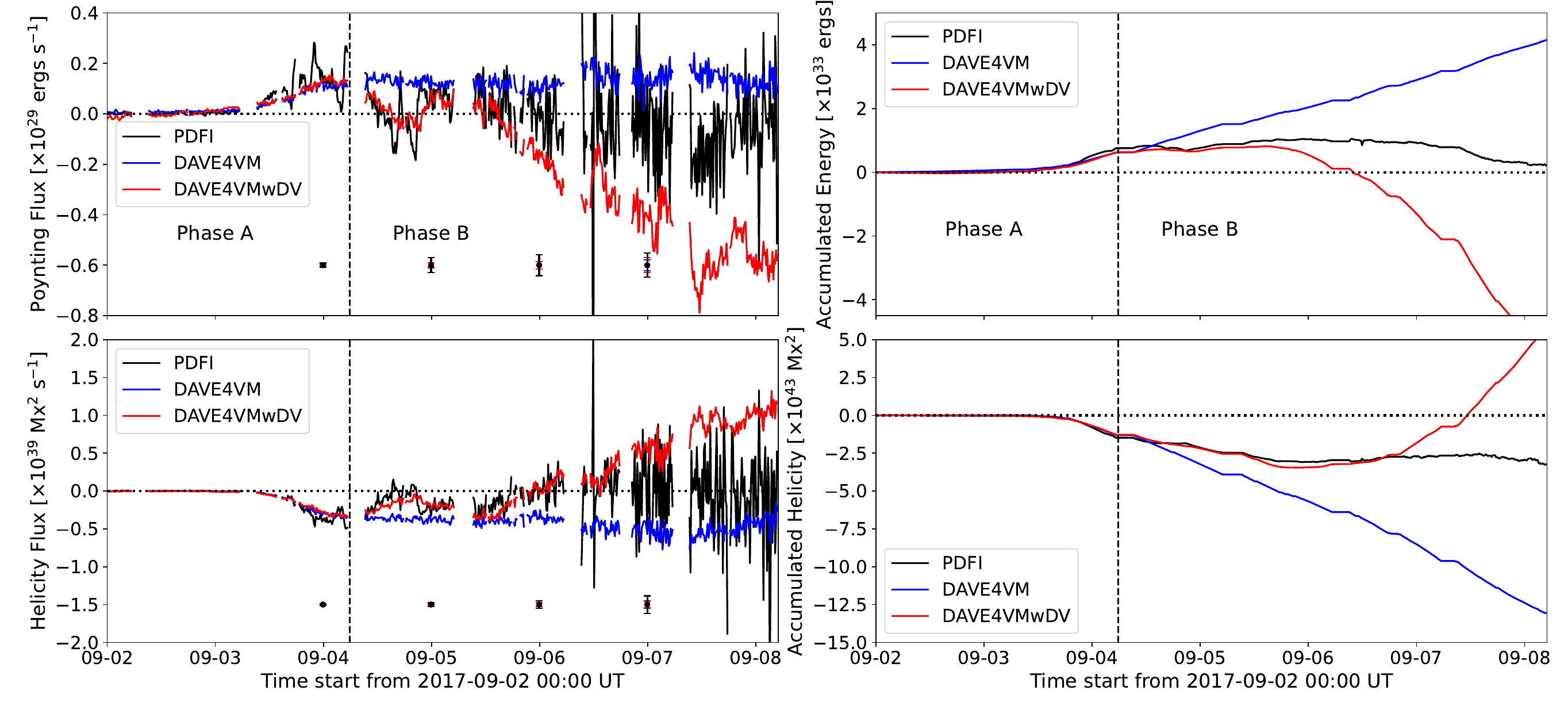}
  \caption{The time evolution of estimated Poynting flux and helicity flux for NOAA AR 12673 from PDFI (black), DAVE4VM (blue), and DAVE4VMwDV (red). Top: The time evolution of Poynting flux (left) and accumulated Poynting flux (right). Here and after, data gaps are assigned \chg{a} zero value for integration. Bottom: The time evolution of helicity flux (left) and accumulated helicity flux (right). Uncertainties due to the noise in HMI magnetograms are shown for four time frames.}
  \label{fig:compare_all}
  \vspace{2mm}
\end{figure*}


\subsection{Calculation of Poynting flux and Helicity flux}

With the inferred $\bm{v}$ or $\bm{E}$, we can calculate the vertical Poynting flux $S_z$ and (relative) helicity flux $dH_m/dt$ \citep{Beger_1984} transport through the photosphere via a surface integral:
\begin{equation}
  \begin{split}
 S_z &= \frac{1}{4\pi}\int_S \bm{E}\times \bm{B} \cdot \hat{\bm{z}} dS \\
  &= \frac{1}{4\pi}\int_S \left[\bm{B}_h^2 v_z - (\bm{B_h}\cdot\bm{v}_h) B_z\right] dS, \\
 \frac{d H_m}{d t} &= -2\int_S \bm{A}_p\times \bm{E} \cdot \hat{\bm{z}} dS \\
  &= -2\int_S \left[(\bm{A}_p \cdot\bm{v}_h) B_z - (\bm{A}_p \cdot \bm{B}_h) v_z \right] dS 
  \end{split}
  \label{equ:Sz_Hm}
\end{equation}
where $S$ denotes the photosphere, and $\bm{A}_p$ is the vector potential for the potential field $\bm{B}_p$ that satisfies 
\begin{equation}
  \begin{split}
    \nabla \times \bm{A}_p \cdot \hat{\bm{z}} &= B_z ,\\
 \bm{A}_p\cdot \hat{\bm{z}} &= 0,
  \end{split}
\end{equation}
and follows the Coulomb gauge
\begin{equation}
  \nabla \cdot \bm{A}_p = 0 .
\end{equation}

As suggested by \cite{schuck_2019}, using the formalism in Equation \ref{equ:de_E}, the Poynting flux contributed by inductive and non-inductive electric fields \chg{is}
\begin{equation}
  \begin{split}
 S_z &= S_z^\text{I} + S_z^\text{NI}, \\ 
 S_z^\text{I} &= \frac{1}{4\pi}\oint_S \bm{E}^\text{I} \times \bm{B} \cdot \hat{\bm{z}} dS = - \frac{1}{4\pi}\oint_S \chi \nabla_h \cdot \bm{B} dS, \\
 S_z^\text{NI} &= \frac{1}{4\pi}\oint_S \bm{E}^\text{NI} \times \bm{B} \cdot \hat{\bm{z}} dS = \frac{1}{4\pi} \oint_S \xi \nabla_h \times \bm{B} dS. \\
  \end{split}
\end{equation}
Similarly, for a closed coronal volume with only photospheric contributions, the helicity flux contributed by inductive and non-inductive electric fields \chg{is}
\begin{equation}
  \begin{split}
 \frac{d H_m}{dt} &= \frac{d H^\text{I}_m}{dt} + \frac{d H^\text{NI}_m}{dt}, \\ 
 \frac{d H^\text{I}_m}{dt} &= -2\oint_S \bm{A}_p\times \bm{E}^\text{I} \cdot \hat{\bm{z}} dS = 0, \\
 \frac{d H^\text{NI}_m}{dt} &= -2\oint_S \bm{A}_p\times \bm{E}^\text{NI} \cdot \hat{\bm{z}} dS = 2\oint_S \xi B_z dS. \\
  \end{split}
  \label{equ:de_Hm}
\end{equation}
It should be noted that the calculation of the helicity flux requires $S$ to be a closed surface, which is usually not true for the application on a specific active region. The calculation of ${d H_m}/{dt}$ for observations on an open surface, as is the case for most magnetogram data, will be discussed in Section~\ref{sec:Inductive}. 

\begin{figure*}[t!]
  \centering
  \includegraphics[width=0.95\textwidth]{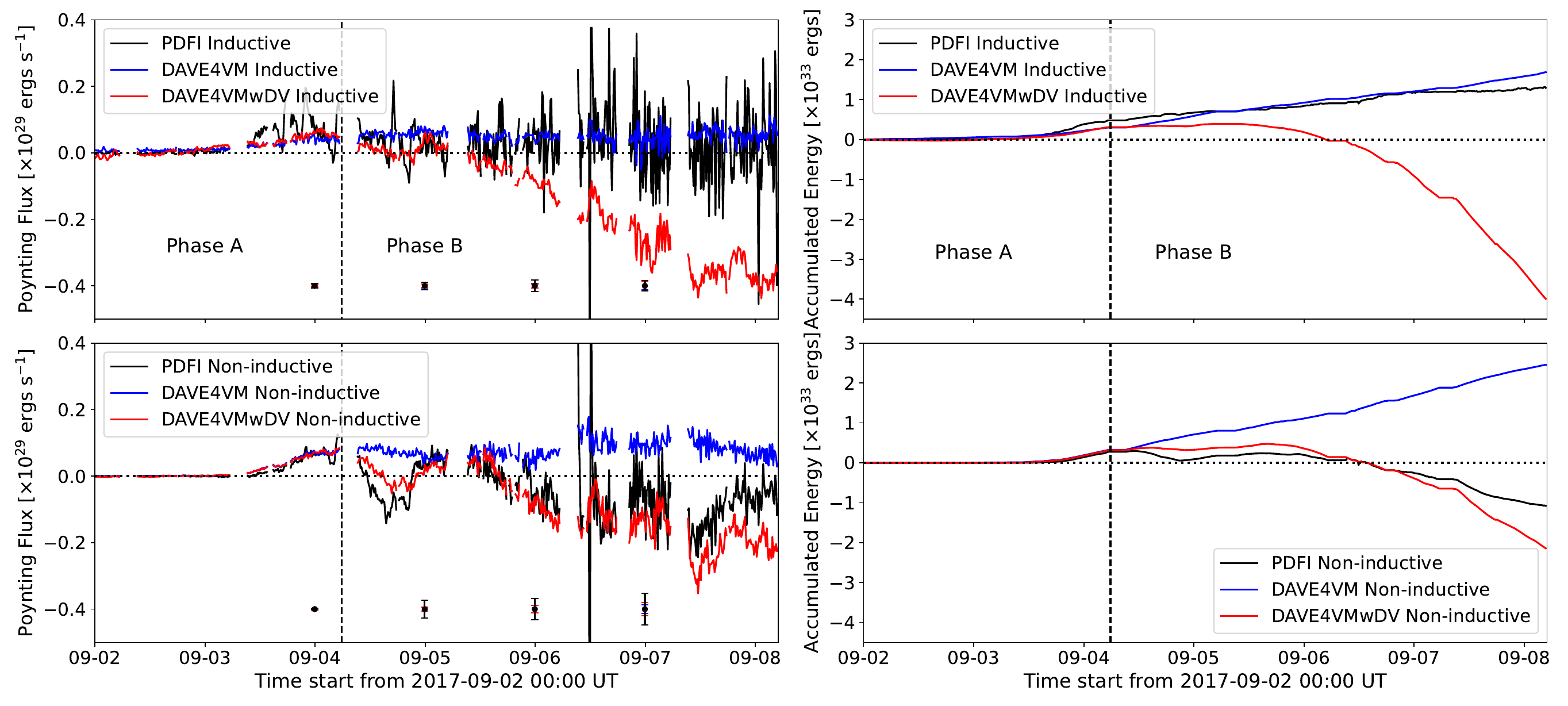}
  \caption{The time evolution of Poynting flux contributed from inductive and non-inductive electric field for NOAA AR 12673. Top: The time evolution of inductive Poynting flux (left) and accumulated inductive Poynting flux (right). Bottom: The time evolution of non-inductive Poynting flux (left) and accumulated non-inductive Poynting flux (right).}
  \label{fig:compare_Sz}
  \vspace{2mm}
\end{figure*}

We use a Monte Carlo method to estimate the uncertainties on Poynting and helicity flux due to the noise in HMI data \chg{\citep{PDFI_2015,Avallone2020,Liu2023}}. We create $N \sim 100$ pairs of magnetograms with different noise realizations centered at September 3rd, 23:54UT, September 4th, 23:54UT, September 5th, 23:54UT, and September 6th, 23:54UT. They are then used as the input of DAVE4VM(wDV) and PDFI for velocity and electric field estimation. After that, the $N$ magnetograms and the velocity or electric field are used to calculate Poynting $S_z$ and helicity flux $ dH_m/ dt$ each time. Finally, we take the standard deviations of the $N$ maps as an estimate of the typical uncertainties, which are typically small (e.g., Figure~\ref{fig:compare_Sz}).


\begin{figure*}[t!]
  \centering
  \includegraphics[width=0.95\textwidth]{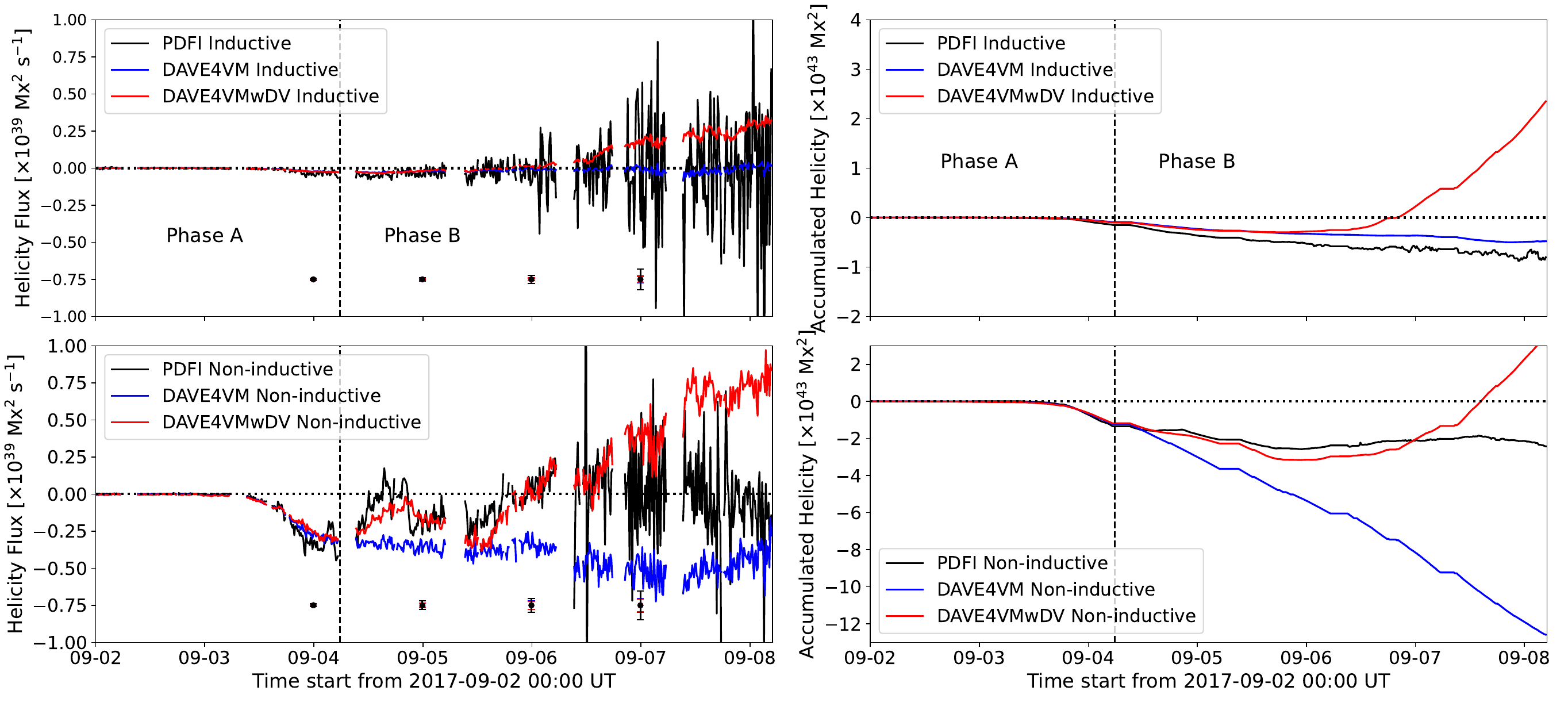}
  \caption{Similar to Figure \ref{fig:compare_Sz}, but for inductive and non-inductive helicity flux.} 
  \label{fig:compare_Hm}
  \vspace{2mm}
\end{figure*}
  

\section{Result}\label{sec:result}

\subsection{Marked Disagreement Between Three Methods}
\label{subsec:discrepancy}

We find that the inferred Poynting and helicity fluxes differ significantly between the three methods. Their temporal evolutions are shown in Figure~\ref{fig:compare_all}, where black, blue, and red curves represent the results from PDFI, DAVE4VM, and DAVE4VMwDV, respectively.

The top-left panel shows the evolution of Poynting flux $S_z$. The three methods agree reasonably well till about September 4, 06:00 UT. Afterwards, the DAVE4VM result diverges from the other two. The PDFI and DAVE4VMwDV results start to diverge drastically from about September 5, 08:48 UT\chg{.} Overall, the PDFI results exhibit stronger temporal variations than the other two methods. The two large oscillations on September 6 are likely due to the transient artifacts in the HMI observations caused by two major flares. The discrepancies appear more striking in the accumulated energy profiles shown in the top-right panel (see also Table~\ref{tab:accumulate}). While the order of magnitude for DAVE4VM and DAVE4VMwDV ($10^{33}$~erg) is reasonable for major flaring ARs, they have opposite signs. The absolute value of PDFI is one order of magnitude smaller. 

The evolution of inferred helicity flux $d H_m/d t$ and accumulated helicity from three methods are shown in the bottom panel of Figure~\ref{fig:compare_all}. Similar to the energy flux, DAVE4VM starts to diverge from the other two on September 4, 05:48UT, whereas \chg{DAVE4VMwDV} starts to diverge on September 5, 08:48UT. The accumulated helicity values from DAVE4VM and DAVE4VMwDV again have opposite signs.

It is interesting to note that the discrepancies between these methods appear to correlate with the increasing magnitude of Doppler velocity (Figure \ref{fig:AR12673_flux}). We will come back to this point later.


\begin{deluxetable}{cq{3.1}q{3.1}q{3.1}}[t!]
  \tablecaption{Accumulated Poynting and helicity flux}\label{tab:accumulate}
  \tablewidth{0pt}
  \tablehead{
  \colhead{\textbf{}} & \colhead{\textbf{PDFI}} & \colhead{\textbf{DAVE4VM}} & \colhead{\textbf{DAVE4VMwDV}}
 }
 \decimals
  \startdata
    $S_z$ & 0.2 & 4.1 & -6.1  \\ 
    $S_z^\text{I}$ & 1.3 & 1.7 & -4.0  \\ 
    $S_z^\text{NI}$ & -1.1 & 2.5 & -2.1 \\ 
    \cline{1-4}
    $dH_m/dt$ & -3.2 & -13.1 & 5.9  \\ 
    $dH_m^\text{I}/dt$ & -0.5 & 1.3 & 2.4   \\ 
    $dH_m^\text{NI}/dt$ & -2.4 & -12.6 & 3.5  \\ 
  \enddata
  \tablecomments{The values are for the end of the studied period, September 8, 05:00UT. The unit for the accumulated Poynting flux is $10^{33}$~erg, and unit for the accumulated helicity flux is $10^{43}$~Mx$^2$.}
  \end{deluxetable}
  

\subsection{Decomposition of Poynting Flux}

The left column of Figure~\ref{fig:compare_Sz} shows the Poynting flux contributed by the inductive electric field $\bm{E}^\text{I}$ and non-inductive electric field $\bm{E}^\text{NI}$. Hereafter, we call the Poynting (helicity) flux contributed from these two terms the inductive and non-inductive Poynting (helicity) flux, and denote the corresponding variables with the superscript `I' and `NI', respectively. Because $\bm{E}^\text{I}$ in PDFI is solved with PTD solution, we denote the inductive Poynting (helicity) flux from PDFI as the PTD results. The accumulated inductive and non-inductive Poynting fluxes are plotted in the right column of Figure~\ref{fig:compare_Sz}; their values at the end of the studied period are listed in the top half of Table~\ref {tab:accumulate}. 

For the inductive Poynting flux, the DAVE4VM values are generally small, whereas the PDFI results oscillate significantly. The accumulated fluxes from these two methods agree relatively well, reaching $1.3\times10^{33}$~erg and $1.7\times10^{33}$~erg, respectively. In comparison, the results from DAVE4VMwDV \chg{deviate} drastically starting around September 5. The final accumulated energy reaches $-4.0\times10^{33}$~erg.

For the non-inductive Poynting flux, the PDFI results appear to agree better with DAVE4VMwDV instead. Both display large negative variations on September 4, \chg{they} continually trend negative after September 5. The accumulated energy reaches $-1.1\times10^{33}$~erg and $-2.1\times10^{33}$~erg, respectively. Meanwhile, the DAVE4VM energy flux stays positive, and the accumulated energy reaches $2.5\times10^{33}$~erg.

We find that the non-inductive Poynting flux contributes significantly to the total energy budget for all three models. The \chg{magnitudes} of the accumulated Poynting flux from the non-inductive and inductive terms are comparable for PDFI but with different signs. The non-inductive portion has the same sign as the inductive counterpart DAVE4VM and DAVE4VMwDV, and \chg{accounts} for $61\%$ and $34\%$ of the total, respectively.


\begin{figure*}[t!]
  \centering
  \includegraphics[width=0.98\textwidth]{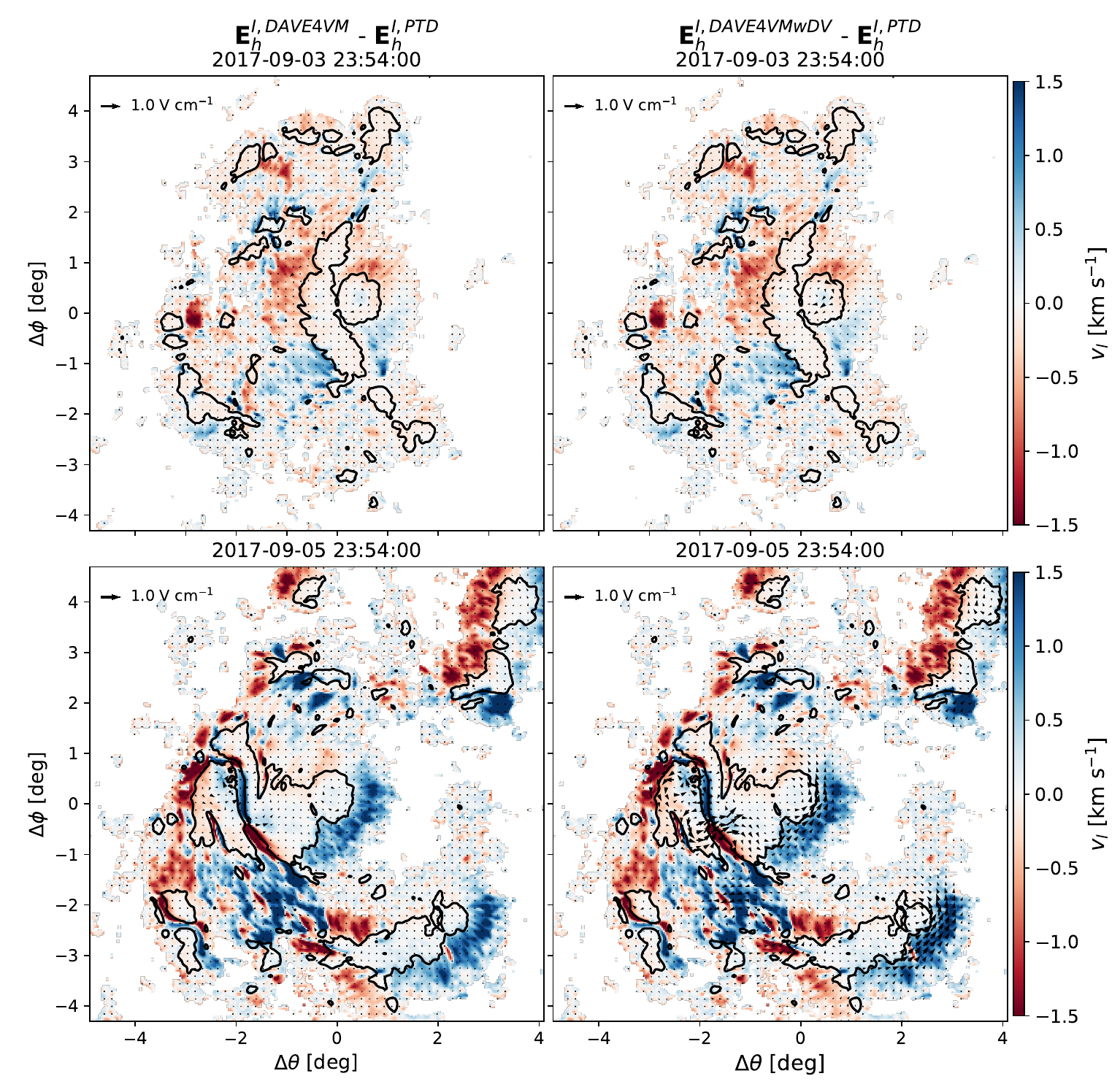}
  \caption{The difference of inductive electric field $\bm{E}^\text{I}$ between PDFI, DAVE4VM, and DAVE4VMwDV \chg{with the background of Doppler velocity $v_l$ map}. Left: The difference in $\bm{E}^\text{I}$ between PDFI and DAVE4VM. Right: The difference in $\bm{E}^\text{I}$ between PDFI and DAVE4VMwDV. The black arrows show the difference of horizontal inductive electric field $\Delta \bm{E}^\text{I}_h$. The background map shows the time derivative of the vertical magnetic field ${\partial B_z}/{\partial t}$. The black contours mark the regions with strong vertical magnetic field $\lvert B_z \rvert > 1000$~G. Here and after, the frames are centered at the center of active region in CGEM dataset. }
  \label{fig:compare_EI}
  \vspace{2mm}
\end{figure*}


\subsection{Decomposition of Helicity Flux}\label{sec:helicity}

Figure~\ref{fig:compare_Hm} shows the temporal evolution of inductive and non-inductive helicity flux on the left, and their accumulation on the right. The bottom half of Table \ref{tab:accumulate} lists the final accumulated inductive and non-inductive helicity flux.

The inductive helicity flux from all three methods are close to zero before September 4. For PDFI, the temporal variation increases drastically after September 5, and the accumulated helicity trends \chg{negatively} to reach $-0.5 \times 10^{43}$~Mx$^2$. The inductive helicity fluxes for DAVE4VM and DAVE4VMwDV show less temporal variation, and the positive trend of the latter during phase B is quite pronounced. Both their accumulated values stay positive, reaching $1.3 \times 10^{43}$~Mx$^2$ and $2.4 \times 10^{43}$ Mx$^2$ respectively. We note that all three values are non-zero. This is a consequence of dealing with an active region patch: for a closed surface, it should be zero \citep{schuck_2019}. We will discuss this point more in Section~\ref{sec:Inductive}.

For the non-inductive helicity flux, all three methods infer a similar amount of negative injection till the end of phase A. The PDFI and DAVE4VMwDV values follow one another closely until September 7, after which the latter again trends positive. Their final accumulated helicity reaches $-2.4 \times 10^{43}$~Mx$^2$ and $3.5 \times 10^{43}$~Mx$^2$, respectively. The DAVE4VM values, on the other hand, \chg{trend negatively}. The accumulated helicity reaches $-12.6 \times 10^{43}$~Mx$^2$.


\section{Discussion}\label{sec:discussion}

We have shown that the inferred Poynting and helicity flux from the three methods are markedly different for AR 12673. The main discrepancy between PDFI and DAVE4VM appears to come from the non-inductive part, while the main discrepancy between PDFI and DAVE4VMwDV appears to come from the inductive part. Below, we investigate the causes and search for techniques to reduce the discrepancy between the solutions.


\begin{figure*}[t!]
  \centering
  \includegraphics[width=0.95\textwidth]{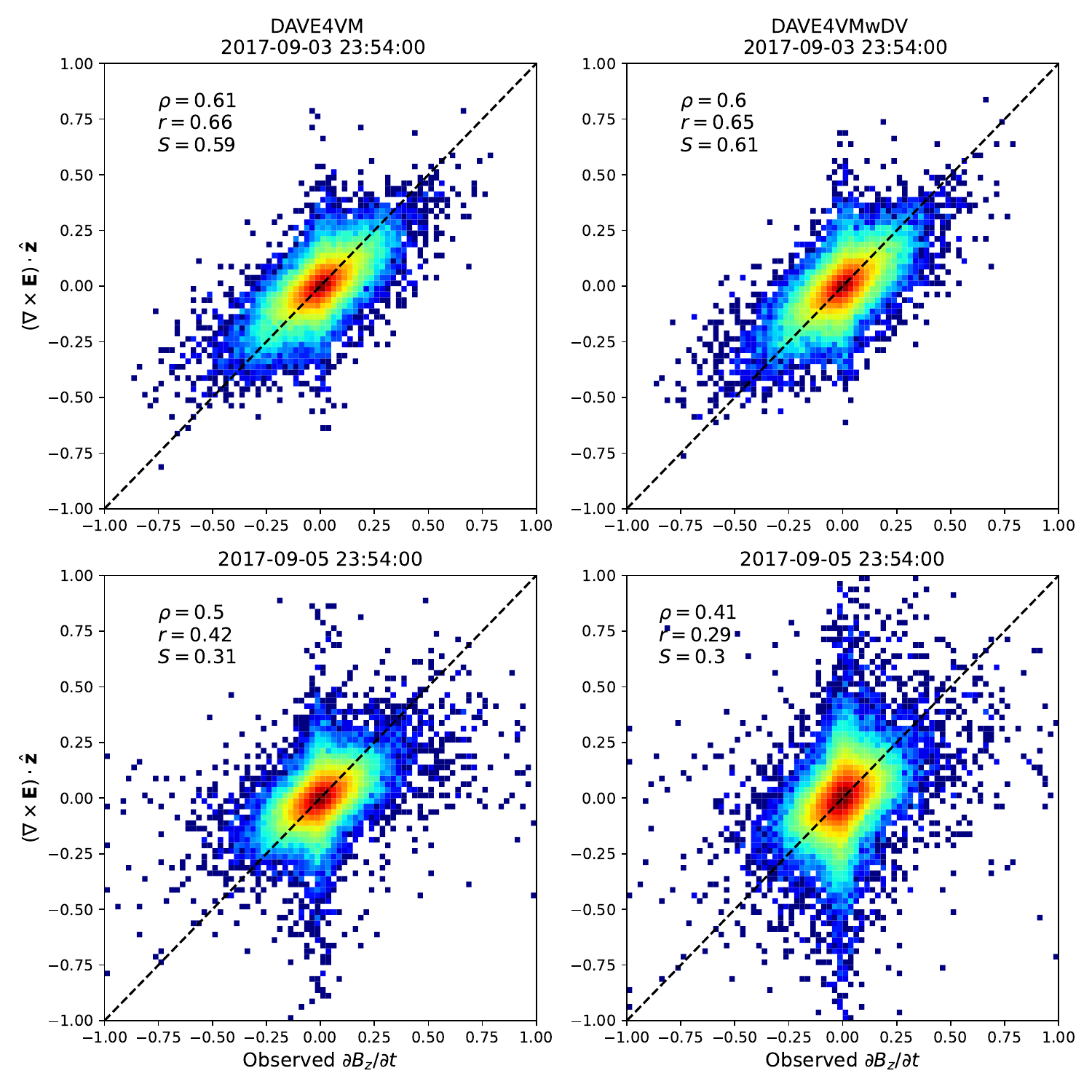}
  \caption{\chg{A comparison between observed $\partial B_z/\partial t$ and the derived $(\nabla\times\bm{E})\cdot \hat{\bm{z}}$ from DAVE4VM (left) and DAVE4VMwDV (right). From top to bottom, we show the comparison at 2017-09-03 23:54:00~UT and 2017-09-05 23:54:00~UT. The Spearman coefficient ($\rho$), Pearson coefficient ($r$), and slope ($S$) are also shown on the plots.} } 
  \label{fig:inductivity}
  \vspace{2mm}
\end{figure*}

\begin{figure*}[t!]
  \centering
  \includegraphics[width=0.95\textwidth]{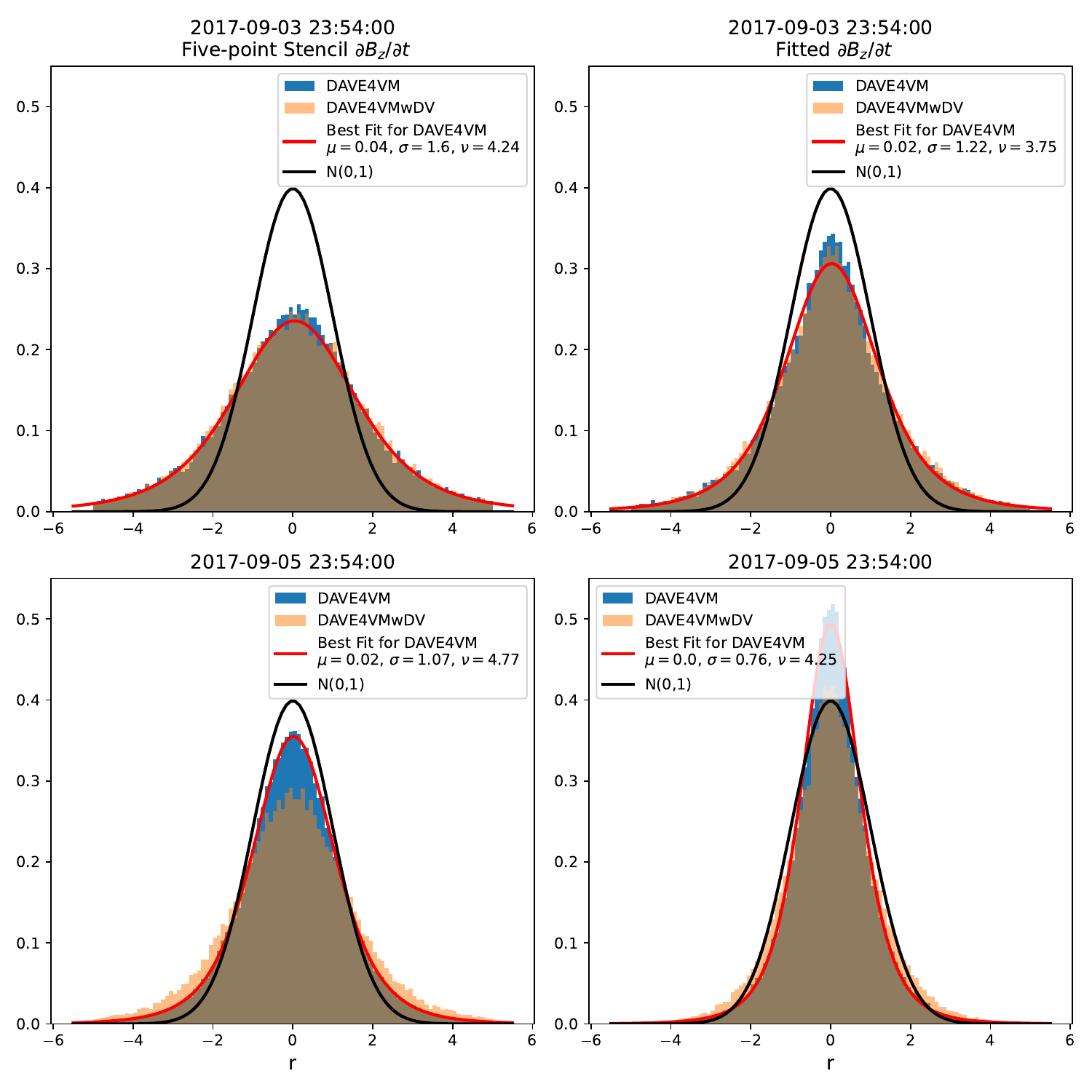}
  \caption{The distributions of $r$, normalized residual of $\partial B_z / \partial t$ (Equation~(\ref{eq:r})) at \chg{September 3, 23:54UT (top) and September 5, 23:54UT (bottom). Left: the distributions of $r$ based on $\partial B_z / \partial t$ calculated with five-point stencil. Right: The distributions for $r$ based on the least-square fit of $\partial B_z / \partial t$.}  A Gaussian distribution $N \sim (0,1)$ is plotted as black curve. The best fit $t$-distribution for DAVE4VM are plotted as red curves. } 
  \label{fig:distribution}
  \vspace{2mm}
\end{figure*}


\subsection{Discrepancies in Inductive Terms}
\label{sec:Inductive} 


\textit{Spatial distribution of discrepancies.}--- We first visualize the differences of the inductive electric field between DAVE4VM and PTD, and the differences between DAVE4VMwDV and PTD in Figure~\ref{fig:compare_EI}. We select September 3, 23:54 UT to represent a time when the three methods agree relatively well, and September 5, 23:54 UT, a time when DAVE4VMwDV diverges from the other two methods. Here we are only interested in the horizontal component of the electric field $\bm{E}^\text{I}_h$, as they solely constrain the vertical component of the induction equation. The differenced $\bm{E}_h$ values are shown as black arrows. The black contours mark the region with $\lvert B_z \rvert > 1000$ G, which is largely co-spatial with the sunspot umbrae. \chg{The regions with large Doppler velocity are co-spatial with penumbrae and regions along certain PILs (cf. Figure~\ref{fig:AR12673}).}

The differenced electric fields, represented by black arrows, are relatively small except in the lower right panel as expected. Differences between DAVE4VM and PDFI (left column) appear to mostly reside in umbral regions. \chg{In the umbrae, $B_z$ is relatively stable, so the genuine $\partial_tB_z$ signals are expected to be small. Random noise may affect the solution more severely there.}

Differences between DAVE4VMwDV and PTD, on the other hand, also extend to regions with large Doppler velocity (red contours). \chg{As discussed below, the inclusion of the Doppler term in DAVE4VMwDV \textit{may} affect the minimization of the inductivity term in Equation~(\ref{l2}), leading to an electric field that is ``less ideal'' and differs from the PTD solution. However, the difference in electric field is not always significant at locations with large $v_l$. When the two terms in Equation~(\ref{l2}) can both be reduced satisfactorily, the solutions from DAVE4VMwDV and PTD will converge.}

In the lower right panel, the differenced electric field shows a counter-clockwise pattern in the core region centered at around coordinate $(-2,0)$. This results in an additional positive contribution of $\nabla \times \bm{E}$, or negative $\partial B_z/\partial t$. 


\begin{figure*}[t!]
  \centering
  \includegraphics[width=0.95\textwidth]{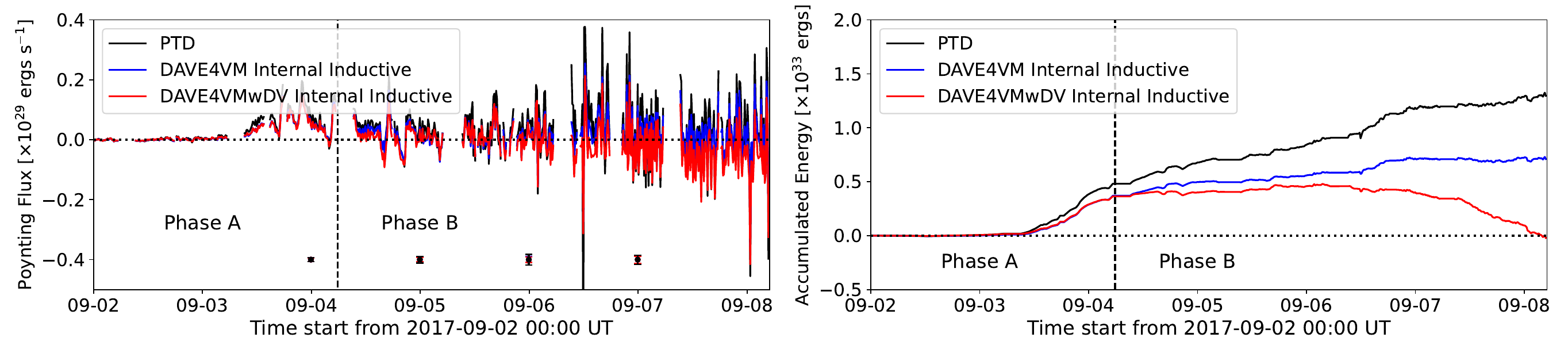}
  \caption{Comparison of $S_z^\text{I}$ and $S_z^\text{NI}$ after adopting the internal fitting. Top: The time evolution of $S_z^\text{I}$ (left) and accumulated $S_z^\text{I}$ (right). Bottom: The time evolution of $S_z^\text{NI}$ (left) and accumulated $S_z^\text{NI}$ (right). Black, blue, and red curves refer to the results from PDFI, DAVE4VM internal fitting, and DAVE4VMwDV internal fitting.} 
  \label{fig:compare_pp}
  \vspace{2mm}
\end{figure*}


\textit{Difference in design philosophy.}--- The small differences between DAVE4VM and PTD may be a consequence of their different design philosophy. While DAVE4VM finds the velocity by fitting the observed $\partial B_z /\partial t$ with the least square method under the constraint of observational uncertainties, PTD solves for the electric field from the induction equation exactly (Equation~\ref{equ:EI}). \chg{The inductivity (how well the induction equation is satisfied) of the electric field from DAVE4VM is assessed in the left column of Figure \ref{fig:inductivity}. As expected for the least square method, the electric field from DAVE4VM does not satisfy the induction equation exactly, and there is a larger deviation when $\partial B_z/\partial t$ is close to $0$. The inductivity from DAVE4VMwDV shown in the right column is less satisfactory than that of DAVE4VM, especially at the time when the Doppler velocity is large.}

\chg{We further} use the residual $r$, defined as the difference between the observed and the fitted $\partial_t B_z$, normalized by observation uncertainty $\sigma_{\partial_t B_z}$, i.e.,
\begin{equation}
 r = \frac{\partial_t B_z^{\text{Obs}} - \partial_t B_z^{\text{DAVE4VM}}}{\sigma_{\partial_t B_z}},
 \label{eq:r}
\end{equation}
to describe the difference of the (curl) of the inductive electric field from DAVE4VM(wDV) and observation. Ideally, because DAVE4VM(wDV) uses least square methods and the measured uncertainties as weighting, the values of $r$ should follow a Gaussian distribution $N\sim(0,1)$.

The \chg{left panels of Figure~\ref{fig:distribution} show the histograms of the residuals between DAVE4VM and observation in blue and the residuals between DAVE4VMwDV and observation in orange. Compared to the Gaussian distribution $N$$\sim$$(0,1)$, the actual distributions of $r$ have a fat tail. The DAVE4VM result can be reasonably described with a $t$-distribution that has a Gaussian core and a fat tail, plotted as a red curve in the figure.}

\chg{One possible reason for the fat tail is that the measured magnetic field is on a constant $\tau$ surface. The issue is that \textit{all} current optical flow methods assume that the $(\tau=1)$ surface is flat or spherical (at a fixed $z$ or $r$) rather than undulating in altitude. To account for this fact that the observations in adjacent pixels vary in altitude, the vertical coordinate implicitly depends on the horizontal coordinates $z=Z(x,y)$ and thus the induction equation for the observations on the $\tau=1$ surface becomes: 
\begin{equation}
 \frac{\partial B_z}{\partial t}=\nabla_h\cdot\left(v_z\bm{B}_h-\bm{v}_{h}B_z\right)+\frac{\partial\left(v_z\bm{B}_h-\bm{v}_{h}B_z\right)}{\partial z}\cdot\nabla_h Z
\end{equation}
So the assumed ideal induction equation model can possibly deviate from the real Sun and result in the non-Gaussian tails, for example, near the boundary of umbra and penumbra where $\nabla_h Z$ is large due to Wilson depression \citep{Wilson}.} Ongoing efforts that provide inversion on a grid of geometric heights \citep[e.g.,][]{FIRTEZ-dz,spin4d2} may alleviate the issue.


\chg{Another possible reason is that as the magnetic field $\bm{B}$ of an active region evolve in time, the relative importance of different spatial scales (i.e., power spectrum) in $\bm{B}$ and $\partial_t \bm{B}$ will also change in time (see discussion on scales below). It is unlikely that the same set of DAVE4VM(wDV) parameters will work well at all stages of an active region's lifetime. Here, the parameters are chosen to optimize performance for the September 5 frame; their performance is worse on the September 3 frame. This result is illustrated by the improved Gaussian fit for the lower left panel of Figure~\ref{fig:distribution} compared to the upper left panel. Parameters that vary spatially and temporally may be needed in order to better reproduce the observed $\partial_t B_z$.}

\textit{Inclusion of Doppler velocity.}--- The distributions of fitted $\partial_t B_z$ and calculated $\partial_t B_z$ from DAVE4VM and DAVE4VMwDV at early times are close to each other. However, the residuals from DAVE4VMwDV exhibit a broader distribution at late time, when the Doppler velocity becomes larger. The worse fit of the induction equation in DAVE4VMwDV may be a consequence of the additional regularization term for Doppler velocity (see Equation~\ref{l2}). \chg{Large Doppler velocity may dominate the loss function $L$ and, as a trade-off for a better fit of the observed $v_l$, cause a worse fit to the induction equation.} The large Doppler velocity, in excess of $1$~km s$^{-1}$, also means that our chosen cadence $720$~s violates the Courant-Friedrichs-Lewy (CFL) condition for HMI's spatial resolution $728$~km. These factors severely limit the capability of DAVE4VMwDV to find a solution that satisfies the induction equation and Doppler velocity simultaneously. \chg{It should be noted that the CFL condition can also potentially affect PDFI in inferring both inductive and non-inductive electric fields.}

\textit{``Internal'' vs ``calculated'' derivatives.}--- The deviation between the observed $\partial _t B_z$ and that from the curl of DAVE4VM(wDV) electric field can increase because of the ``offset'' between the independent windows used to infer $\bm{v}$ at neighboring pixels. As mentioned in Section~\ref{sec:methods}, the derivative of $\bm$ is calculated using Equation~(\ref{equ:internal}) \chg{internally in the least square procedure}. However, in practice, the derivative of $\bm{v}$ is often calculated using finite difference method that involves $\bm{v}$ from different windows, the derivative of velocity $\bm{v}$ then becomes
\begin{equation}
 \frac{\partial \bm{v}}{\partial x_k} = \sum_{i=0}^d \sum_{j=0}^{d-i}\frac{\partial\bm{A}_{i,j}(x,y)}{\partial x_k} P_i(x_0) P_j(y_0),
  \label{equ:tiles}
\end{equation}
where $x_0$ and $y_0$ are the coordinates of the window center. This difference can result in further discrepancies, which is illustrated in the \chg{right} panels of Figure \ref{fig:distribution}. As we use a five-point stencil to calculate the derivative instead of using the internal derivative, the residual increases. Hereafter, we call $\partial_t B_z$ calculated from five-point stencil as the ``calculated'' version, as \chg{opposed} to the fitted version provided by DAVE4VM(wDV).

Here we re-calculate $\bm{E}^\text{I}$ \chg{from DAVE4VM(wDV) using the ``internal derivatives'' (Equation~(\ref{equ:internal}))} and demonstrate below its effect in reducing the discrepancies.

\chg{We solve for $\bm{E}^\text{I}$ with Equation~(\ref{equ:solution}), but opt to solve scalar $\chi$ with the \textit{fitted} $\partial B_z^\text{LS}/\partial t$ , which is obtained with the velocity and its derivatives (Equation \ref{equ:internal}), i.e.}:
\begin{equation}
  \label{equ:postprocess}
  \begin{split}
  \nabla_h^2 \chi &= \frac{\partial B_z^\text{LS}}{\partial t}. \\ 
  \end{split}
\end{equation}
The corrected $S_z^\text{I}$ values from the three methods are shown in Figure \ref{fig:compare_pp}. Compared to the original results in Figure~\ref{fig:compare_Sz}, the corrected $S_z^\text{I}$ values from DAVE4VMwDV now agree better with PDFI, though with larger temporal fluctuations. The disagreement between PDFI and DAVE4VM also appears to increase slightly (note the different scale between Figures~\ref{fig:compare_pp} and \ref{fig:compare_Sz}). Nevertheless, the accumulated $S_z^\text{I}$ values from all three methods now have the same, positive sign. 


\begin{figure*}[t!]
  \centering
  \includegraphics[width=0.95\textwidth]{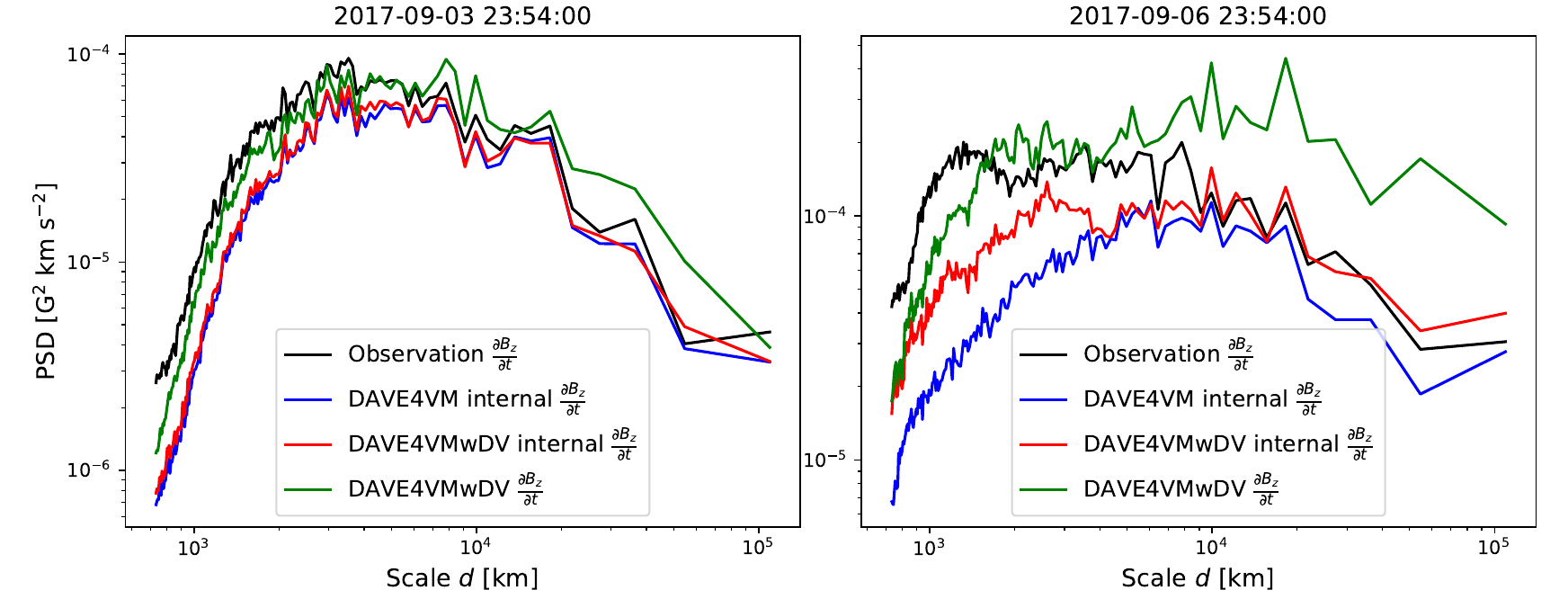}
  \caption{PSD for selected variables as a function of spatial scale $d$ at September 3, 23:54 UT (left) and September 6, 23:54 UT (right). The black, blue, and red lines represent the time derivative of the vertical magnetic field from observation, internal fitting of DAVE4VM and DAVE4VMwDV. The green line represents the time derivative of the vertical magnetic field from DAVE4VMwDV calculated from velocity.} 
  \label{fig:PSD}
  \vspace{2mm}
\end{figure*}


\textit{Dependence on spatial scales.}--- The difference among three methods may be the result of missing small scale structure in $\partial_t B_z$. The least-square nature of DAVE4VM(wDV) and the relatively low order of the Legendre polynomials for fitting \chg{imply} that it is relatively insensitive to noise. However, the smoothness of the solution may also fail to capture small-scale structures in the observations. To investigate the consequence, we show the power spectrum density (PSD) of the fitted $\partial B_z^{\text{LS}}/\partial t$ from DAVE4VM (blue) and DAVE4VMwDv (red) in the top panel of Figure~\ref{fig:PSD}. We also plot the values derived from the observation (black), and those from finite difference (green). For both times, the power of $\partial B_z/\partial t$ in small scale is underestimated, while large scale \chg{agrees relatively} well. The figure also shows that the finite difference results agree much worse on all scales.

\textit{A note on helicity flux.}--- The helicity flux is physically meaningful only for a closed surface. \cite{schuck_2019} performed theoretical analysis and proved that the net inductive helicity flux for a closed surface is zero. However, our results in Section \ref{sec:helicity} show a non-zero net inductive helicity flux. This disagreement comes from the limitation of observation that only covers a portion of solar surface. This limitation can also affect the estimate on non-inductive helicity flux. Here, we show that the non-zero helicity flux is expected when performing integration over a non-closed surface where the net magnetic flux is not zero.

Using the expression of $\bm{E}^\text{I}$ and $\bm{E}^\text{NI}$ in Equation~(\ref{equ:de_E}), ${dH_m^\text{I}}/{dt}$ and ${dH_m^\text{NI}}/{dt}$ accross the surface $S$ in Equation~(\ref{equ:de_Hm}) become 
\begin{equation}
  \begin{split}
 \frac{d H_m^\text{I}}{dt} &= -2\oint_S \mathbf{A}_p \times \nabla \times (\chi\hat{\mathbf{z}}) \cdot \hat{\mathbf{z}} d S \\
  &= 2\oint_{\partial S} \mathbf{A}_p \cdot \mathbf{n} \chi dl,
  \end{split}
\end{equation}
and 
\begin{equation}
  \begin{split}
 \frac{d H_m^\text{NI}}{dt} &=  2\oint_S \mathbf{A}_p \times \nabla \xi \cdot \hat{\mathbf{z}} d S \\
  &= 2\oint_S \xi B_z dS - 2\oint_{\partial S} \mathbf{A}_p \times \mathbf{n} \xi dl,
  \end{split}
\end{equation}
\chg{where $\bm{n}$ is normal vector of the surface $S$, and $\partial S$ is the boundary of surface $S$. Compared to Equation~(\ref{equ:de_Hm}), two additional terms emerge that are related to the boundary condition of the surface}:
\begin{equation}
  \begin{split}
  \oint_{\partial S} \mathbf{A}_p \cdot \mathbf{n} \chi dl,\\
  \oint_{\partial S} \mathbf{A}_p \times \mathbf{n} \xi dl.
  \end{split}
  \label{equ:sub}
\end{equation} 

Under closed surface conditions, these terms are zero, and the results agree with \cite{schuck_2019}. However, in real observation, the field of view is usually a portion of the solar \chg{photosphere resulting in an open surface}, which violates the conditions required for Equation~(\ref{equ:de_Hm}). 


\begin{figure*}[t!]
  \centering
  \includegraphics[width=0.95\textwidth]{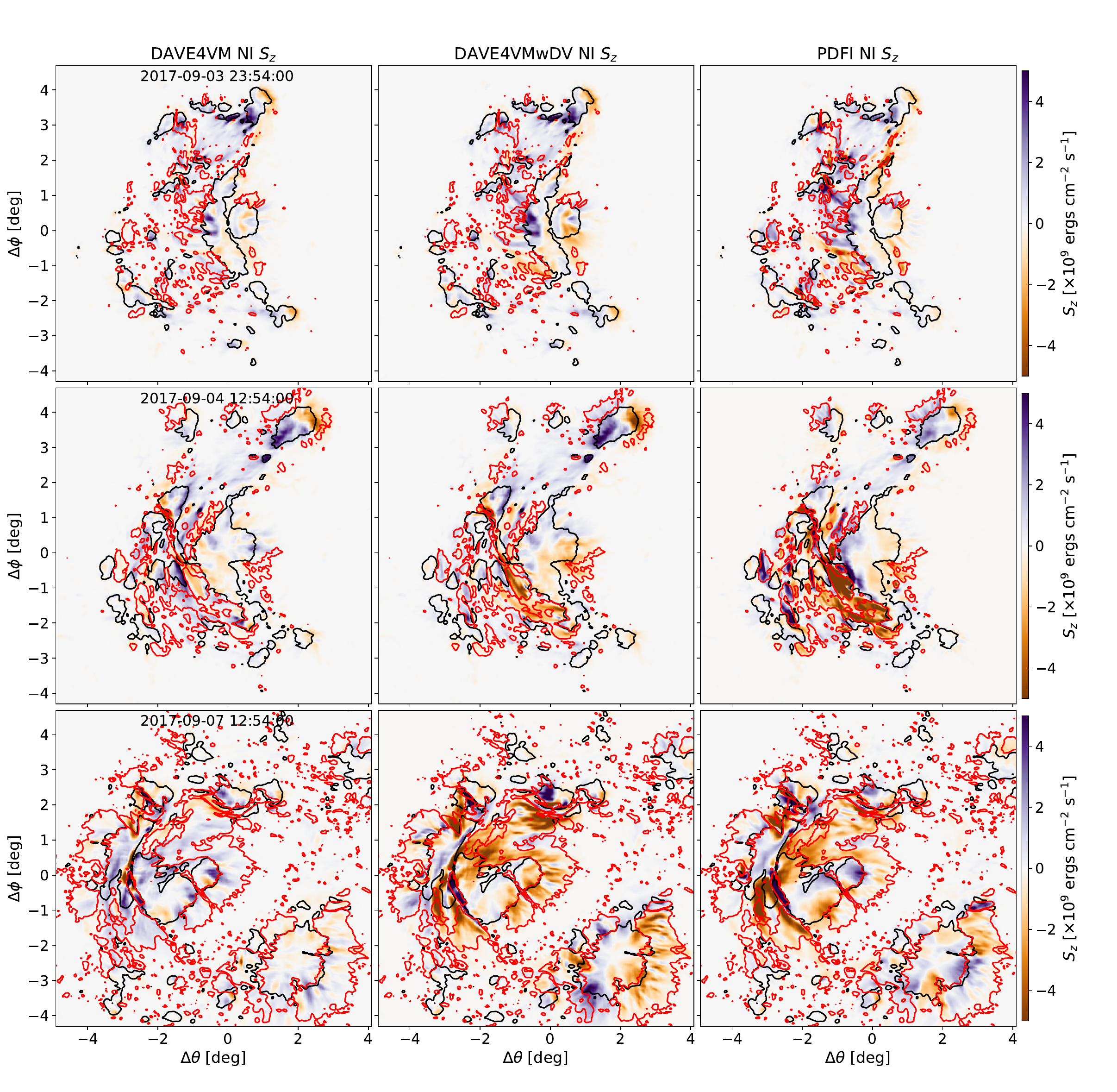}
  \caption{Comparison of non-inductive Poynting flux distributions. From left to right, we show the $S_z^\text{NI}$ maps from DAVE4VM, DAVE4VMwDV, and PDFI, respectively. The top, middle, and bottom \chg{rows} are for September 3 23:54~UT, September 4 23:54~UT, and September 7 12:54~UT, respectively. The red contours mark the regions with Doppler velocity $\lvert v_l \rvert > 0.5$ km s$^{-1}$.}
  \label{fig:compare_NI}
\end{figure*}


\chg{The two additional terms in Equation~(\ref{equ:sub}) are arbitrary in the sense that they depend on the specific choice of the boundary if only the helicity injection through the photosphere is considered. More importantly, they are not expected to meaningfully contribute to the helicity injection when considering a box with one face in the photosphere. This is because the line integrals along any edge of the photosphere will be exactly canceled by the line integrals along the same edge associated with the side faces.}

One way to remove the effect of first term is to use the following equation instead \citep{Beger_1984,schuck_2019}:
\begin{equation}
 \frac{d H_m}{dt} =  -2\int_S \mathbf{A}_p \times (\bm{E} - \bm{E}_p) \cdot \hat{\mathbf{z}} d S
\end{equation}
where $\bm{E}_{p}$ is the electric field that drives the evolution of the potential field, and its horizontal components are identical to that of the inductive electric field, $\bm{E}_{p,h} = \bm{E}_h^\text{I}$. \chg{The second term can be removed by adopting the following Dirichlet boundary: $\xi=0$ for $\partial S$. In fact, this is the exact boundary condition used to solve} for $\bm{E}^\text{NI}=\nabla_h\xi$ (see Equations~(\ref{equ:boundary}) and (\ref{equ:de_E})). This implies that the corrected helicity flux is simply the non-inductive helicity flux shown in the bottom panel of Figure~\ref{fig:compare_Hm}.


\begin{figure*}[t!]
  \centering
  \includegraphics[width=0.95\textwidth]{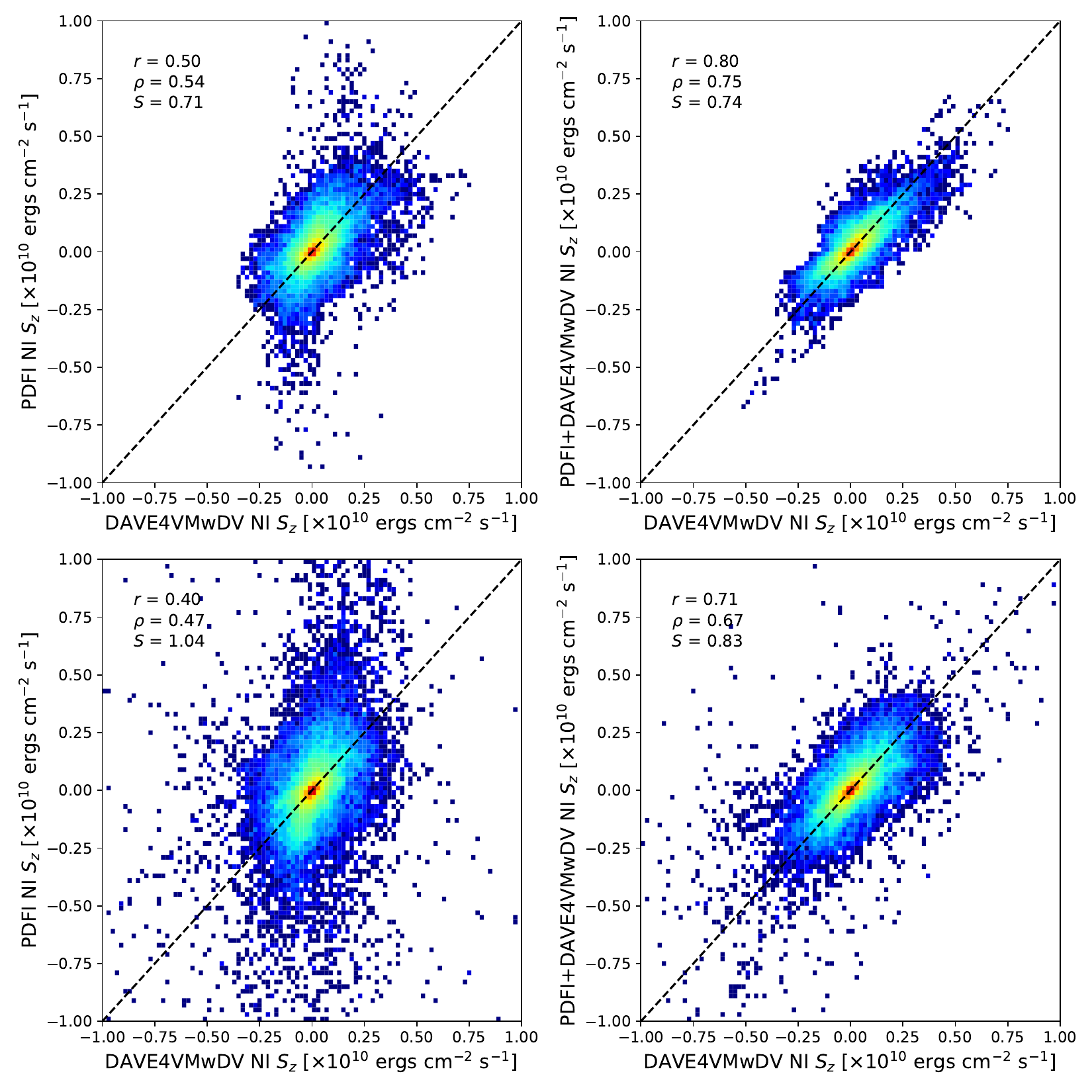}
  \caption{The two-dimension (2D) histograms of non-inductive Poynting flux $S_z^\text{NI}$ from different methods. From left to right, we show the 2D histogram of $S_z^\text{NI}$ between DAVE4VMwDV and PDFI, and DAVE4VMwDV and PDFI+DAVE4VMwDV. The top panel shows the comparison on September 3, 23:54UT. The bottom panel shows the comparison on September 4, 23:54UT. The Pearson coefficient ($r$), Spearman coefficient ($\rho$), and the slope $S$ of the linear fitting are shown on the 2D histograms.} 
  \label{fig:hist_NI}
  \vspace{2mm}
\end{figure*}

\begin{figure*}[t!]
  \centering
  \includegraphics[width=0.95\textwidth]{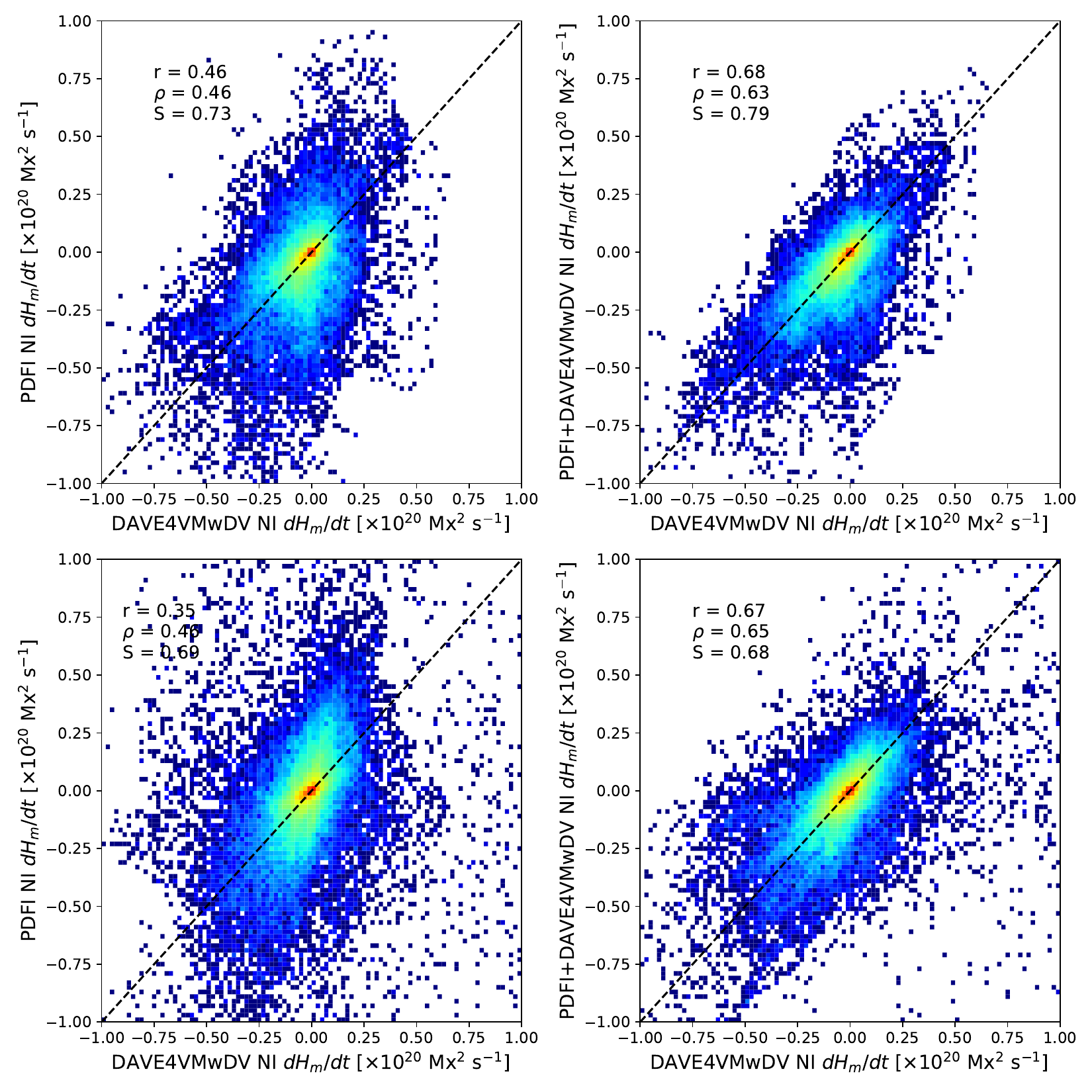}
  \caption{Similar to Figure \ref{fig:hist_NI} but for non-inductive helicity flux.} 
  \label{fig:hist_NI_Hm}
  \vspace{2mm}
\end{figure*}


\subsection{Discrepancies in Non-Inductive Terms}
\label{sec:NonInductive} 

\textit{Effect of Doppler velocity.}--- To investigate the reason for the discrepancy in the non-inductive contributions, we study three representative frames: September 3 23:54~UT,  September 4 23:54~UT, and September 7 12:54~UT. They exhibit quite different Doppler signals as well as non-inductive contributions. The detailed maps of $S_z^\text{NI}$ are shown in Figure \ref{fig:compare_NI}. Similar to Figure~\ref{fig:compare_EI}, regions of strong $B_z$ and large $|v_l|$ are marked with black and red contours, respectively.

One main difference between DAVE4VM and the other two methods is that it does not use Doppler data as input. This directly leads to large disagreement in $S_z^\text{NI}$ in the region with prominent Doppler velocity. For example, on September 4 when the central $\delta$-spot started to form with high Doppler velocity (middle row of Figure~\ref{fig:compare_NI}), the PDFI map show only moderate negative $S_z^\text{NI}$ values along the PIL centered around $(-1.8^\circ, -1.0^\circ)$, whereas the other two methods show much stronger negative $S_z^\text{NI}$ values. Consequently, PDFI infers the most positive Poynting flux among the three methods. In contrast, the September 3 frame \chg{exhibits} weaker Doppler signals (top row of Figure~\ref{fig:compare_NI}); the $S_z^\text{NI}$ maps and the integrated values are much similar. These results suggest that Doppler velocity can significantly affect the inferred non-inductive electric field and the Poynting flux, as readily pointed out by \citet{Fisher_2012}.

As discussed in Section~\ref{subsec:discrepancy}, the discrepancies of the accumulated non-inductive Poynting flux \chg{continue} to increase, and \chg{appear} to accelerate after September 6 (Figure~\ref{fig:compare_Sz}). Inspection of the $S_z^\text{NI}$ maps at late times suggest significant contribution from the penumbral regions (bottom row of Figure~\ref{fig:compare_NI}), which may not be real as the Evershed flows in penumbral regions are field-aligned flows, which should not produce large Poynting flux. This result suggests the necessity to revisit how the Doppler velocity should be incorporated in DAVE4VMwDV: a spatially dependent approach is probably advisable by carefully adjusting the weighting $\omega_{v_l}$ according to the local conditions. For PDFI, the Doppler observation is given little weight away from the PILs based on the LOS field (Equation~\ref{equ:PIL_weighting}). However, apparent PILs may appear in \chg{the} penumbral region due to \chg{the} projection effect, and induce undesired large weighting for the Evershed flow signals (B. T. Welsch, private communication). 

\textit{Difference in algorithm designs.}--- The non-inductive contributions of PDFI and DAVE4VMwDV differ in a couple \chg{of} ways, despite that both use Doppler data as input. First, they treat the velocity data differently. PDFI emphasizes the Doppler (FLCT) velocities near (away from) the PIL regions using an ad hoc weighting (Equation~\ref{equ:PIL_weighting}), whereas we adopt a weighting scheme based on the uncertainty of Doppler velocity for DAVE4VMwDV in this work. Second, they enforce the ideality of the electric field ($\bm{E} \cdot \bm{B} = 0$) differently. For PDFI, a correctional scalar $\psi^C$ is calculated, and its gradient \chg{is} appended to the non-inductive electric field to ensure that the final $\bm{E}$ satisfies the ideal condition (see Equation~(\ref{equ:E})). On the other hand, no correction is needed for DAVE4VMwDV as it intrinsically satisfies the ideal condition (see Equation~(\ref{equ:ohm})).

To quantitatively illustrate the effect of these different treatments, we conduct the following experiment termed as ``PDFI+DAVE4VMwDV.'' We replace the input FLCT horizontal velocity and observed Doppler velocity in the PDFI pipeline with the horizontal velocity and fitted Doppler velocity from DAVE4VMwDV. i.e., setting
\begin{equation}
  \begin{split}
 \bm{v}_h^{\text{FLCT}} = \bm{v}_h^{\text{DAVE4VMwDV}}, \\
 v_l = \hat{\bm{\eta}} \cdot \bm{v}^{\text{DAVE4VMwDV}},
  \end{split}
\end{equation}
in Equation~(\ref{equ:PDFI_pipeline}). We show the results in Figure \ref{fig:hist_NI}. Compared to PDFI, the results from PDFI + DAVE4VMwDV have better agreement with DAVE4VMwDV. The Pearson coefficient is increased from $0.50$ to $0.80$ for the September 3 frame. and increased from $0.40$ to $0.71$ for the September 4. The results suggest that a significant fraction of the difference between PDFI and DAVE4VMwDV, in \chg{terms} of the non-inductive contribution, comes from \chg{the different approaches they adopt to infer and to incorporate the velocity field information}. 

We have performed a similar test for non-inductive helicity flux and the results are shown in Figure \ref{fig:hist_NI_Hm}. Similar to the result for non-inductive Poynting flux, the results for non-inductive helicity flux from PDFI + DAVE4VMwDV have better agreement with DAVE4VMwDV. The Pearson coefficient is increased from $0.46$ to $0.68$ for the September 3 frame. and increased from $0.35$ to $0.67$ for the September 4.


\section{Conclusion}\label{sec:conclusion}

In this work, we apply PDFI, DAVE4VM, and DAVE4VMwDV on active region 12673 and compare the Poynting flux and helicity flux inferred from them. To address the reasons for the discrepancies among the results from these methods, we then decompose the inverted electric field into an inductive part and a non-inductive part and investigate their contribution to the Poynting flux and helicity. Our main findings are as follows:
\begin{enumerate}
  \item The inferred Poynting and helicity flux from the three methods have large discrepancies. The values of the accumulated energy and helicity differ significantly between the three methods even in signs. 
  \item \chg{Using Helmholtz-Hodge decomposition, we show that the main discrepancy between PDFI and DAVE4VM comes from the non-inductive contributions, which occur because DAVE4VM omits Doppler velocity information. Conversely, the main discrepancy with DAVE4VMwDV arises from the inductive part, due to DAVE4VMwDV cannot simultaneously satisfy the induction equation and the Doppler velocity when Doppler velocity is large.}
  \item The discrepancy between different estimates of the inductive Poynting flux comes from the offset between velocities among tiles and the trade-off of including Doppler velocity. It also comes from the fact that the difference of design philosophy behind PDFI and DAVE4VM(wDV): PDFI solves the electric field from the induction equation exactly, while DAVE4VM(wDV) inverts the velocity field by fitting the induction equation with a least square method. 
  \item \chg{The discrepancy in the inductive Poynting fluxes can be reduced by adopting the least square fitting of $\partial B_z/\partial t$ that is calculated with base function (Legendre polynomials) provided by DAVE4VM(wDV), rather than differencing the final electric field $\bm{E}$ calculated with output $\bm{v}$ map and $\bm{B}$ map.} 
  \item Doppler velocity can contribute significantly to the non-inductive Poynting flux and helicity flux and thus should be used as input. Different \chg{ways of implementing velocity} may bring different results in estimating non-inductive Poynting flux and helicity flux. A proper constraint on non-inductive electric field requires further investigation.  
  \item To properly estimate the helicity flux on the active region, the equation 
  \begin{equation}
 \frac{d H_m}{dt} =  -2\int_S \mathbf{A}_p \times (\bm{E} - \bm{E}_p) \cdot \hat{\mathbf{z}} d S
\end{equation}
 should be used. 
\end{enumerate}

Estimating Poynting flux and helicity flux based on the photospheric observation is difficult. Despite including Doppler velocity in the inversion pipeline, inverting the key quantity, electric field, or velocity field from PDFI or DAVE4VMwDV is still an ill-posed problem and requires more constraints. Besides that, the observation cadence is also a factor for the CFL condition that makes the inverted velocity or electric field satisfy the induction and Doppler velocity simultaneously. The challenges in constraints and cadence can be partially solved by combining observations from other instruments. The Polarimetric and Helioseismic Imager \citep[PHI,][]{PHI} onboard the Solar Orbiter \citep{SolarOrbiter1,SolarOrbiter2} can provide Doppler velocity observed from a different direction than HMI. The Doppler velocity from a different direction can be an important constraint for DAVE4VMwDV. The Visible Tunable Filter (VTF) on Daniel K. Inouye Solar Telescope \citep[DKIST,]{DKIST,DKSIT_PLAN} and Near Infrared Imaging Spectropolarimeter \citep[NIRIS,][]{NIRIS} on board the Goode Solar Telescope \citep[GST,][]{GST} can observe active \chg{regions} with shorter cadence and higher spatial and spectral resolution than HMI. Their observation can help study the CFL condition in DAVE4VM(wDV) and PDFI. 
Diffraction Limited Near Infrared Spectropolarimeter \citep[DL-NIRSP,][]{DLNIRSP} on board the DKIST performs high spatial, spectral, and temporal observation and helps measure the magnetic field and Doppler velocity at multiple heights simultaneously, although the small FOV limits the observation on quiet Sun. The height information may provide constraints on all three components of the induction equation. \\
\\We thank the ISSI Flow-tracking team (led by M. D. Kazachenko and B. Tremblay), in particular B. T. Welsch, for helpful discussions. Support for this work comes from the National Science Foundation through the \textit{DKIST} Ambassadors program, administered by the National Solar Observatory and the Association of Universities for Research in Astronomy, Inc. X.S. is additionally supported by NSF award 1848250 and NASA award 80NSSC20K1283. The \textit{SDO} data are courtesy of NASA and the HMI science teams.

\bibliography{reference}{}
\bibliographystyle{aasjournal}

\end{CJK*}
\end{document}